\documentclass[useAMS,usenatbib,usegraphicx]{mn2e}
\usepackage{graphicx}
\usepackage{color}
\usepackage{rotating}

\def\mgb{Mg$\,b$}
\def\hb{H$\beta$}
\def\galex{{\it GALEX}}
\def\sauron{{\tt SAURON}}
\begin{document}
%
%
\title[The SAURON project - XXI]{The SAURON project-XXI. The spatially-resolved UV--line strength relations of early-type galaxies}
\author[Jeong et al.]{Hyunjin Jeong,$^{1}$\thanks{E-mail:hyunjin@kasi.re.kr} Sukyoung K. Yi,$^{2,3}$ Martin Bureau,$^{4}$ Roger L. Davies,$^{4}$
\newauthor Roland Bacon,$^{5}$ Michele Cappellari,$^{4}$ P. Tim de Zeeuw,$^{6,7}$ Eric Emsellem,$^{5,6}$
\newauthor Jes\'{u}s Falc\'{o}n-Barroso,$^{8,9}$ Davor Krajnovi\'{c},$^{6}$ Harald Kuntschner,$^{6}$
\newauthor  Richard M. McDermid,$^{10}$ Reynier F. Peletier,$^{11}$ Marc Sarzi,$^{12}$
\newauthor Remco C. E. van den Bosch$^{13}$ and Glenn van de Ven$^{13}$ \\
$^1$Korea Astronomy and Space Science Institute, Daejeon 305-348, Korea\\
$^2$Department of Astronomy, Yonsei University, Seoul 120-749, Korea\\
$^3$Yonsei University Observatory, Seoul 120-749, Korea\\
$^4$Sub-Department of Astrophysics, University of Oxford, Denys
Wilkinson Building, Keble Road, Oxford OX1 3RH\\
$^5$Universit\'{e} de Lyon 1, CRAL, Observatoire de Lyon, 9
av. Charles Andr\'{e}, F-69230 Saint-Genis Laval; CNRS, UMR 5574,\\ ENS de Lyon, France\\
$^6$European Souther Observatory, Karl-Schwarzchild-Str. 2, 85748, Garching, Germany\\
$^7$Leiden Observatory, Leiden University, Niels Bohrweg 2, 2333 CA Leiden, the Netherlands\\
$^8$Instituto de Astrof\'{i}sica de Canarias, V\'{i}a L\'{a}ctea s/n, La Laguna, Tenerife, Spain\\
$^9$Departamento de Astrof\'{i}sica, Universidad de La Laguna (ULL), E-38205 La Laguna, Tenerife, Spain\\
$^{10}$Gemini Observatory, 670 North A'Ohoku Place, Hilo, Hawaii 96720, U.S.A.\\
$^{11}$Kapteyn Astronomical Institute, University of Groningen, P.O. Box 800, 9700 AV Groningen, the Netherlands\\
$^{12}$Centre for Astrophysics Research, University of Hertfordshire, Hatfield AL10 9AB\\
$^{13}$Max Planck Institute for Astronomy, D-69117 Heidelberg, Germany}
\maketitle
%
%
\begin{abstract}
The unexpected rising flux of early-type galaxies at decreasing
ultraviolet (UV) wavelengths is a long-standing mystery. One
important observational constraint is the correlation between
UV$-$optical colours and Mg$_{2}$ line strengths found by
\citeauthor*{bbbfl88} (\citeyear{bbbfl88}). The simplest
interpretation of this phenomenon is that the UV strength is related
to the Mg line strength. Under this assumption, we expect galaxies
with larger Mg gradients to have larger UV colour gradients. By
combining UV imaging from \galex, optical imaging from MDM and
\sauron\ integral-field spectroscopy, we investigate the
spatially-resolved relationships between UV colours and stellar
population properties of 34 early-type galaxies from the \sauron\
survey sample. We find that galaxies with old stellar populations
show tight correlations between the FUV colours (FUV$-V$ and
FUV$-$NUV) and the \mgb\ index, \hb\ index and metallicity [Z/H].
The equivalent correlations for the Fe5015 index, $\alpha$-enhancement
[$\alpha$/Fe] and age are present but weaker. We have also
derived logarithmic internal radial colour, {\it measured} line
strength and {\it derived} stellar population gradients for each
galaxy and again found a strong dependence of the FUV$-V$ and
FUV$-$NUV colour gradients on both the \mgb\ line strength and the
metallicity gradients for galaxies with old stellar populations. In
particular, global gradients of \mgb\ and [Z/H] with respect to the
UV colour (e.g. $\Delta$(\mgb)/$\Delta$(FUV$-$NUV) and
$\Delta$[Z/H]/$\Delta$(FUV$-$NUV)) across galaxies are consistent
with their local gradients within galaxies, suggesting that the
global correlations also hold locally. From a simple model based on
multi-band colour fits of UV upturn and UV-weak galaxies, we have
identified a plausible range of parameters that reproduces the
observed radial colour profiles. In these models, the centers of
elliptical galaxies, where the UV flux is strong, are enhanced in
metals by roughly 60\% compared to UV-weak regions.

\end{abstract}
%
%
\section{INTRODUCTION}
\label{sec:intro}
Far ultraviolet (FUV) radiation was first detected in early-type
galaxies (NGC\,1291, 1316, 1553, 4406, 4486, 4649 and the bulge of
M31) by the {\it Orbiting Astronomical Observatory-2} in 1969
\citep*{c69, cwp72}. This discovery was surprising, since it had
been expected that such old populations would be entirely dark in
the FUV. Even more surprisingly, a few elliptical galaxies showed
extended {\em rising} flux for decreasing wavelengths below
$\approx2000$~\AA\ \citep[e.g.][]{cw79,bcho80}. This was called the
UV upturn phenomenon.

Early-on, one of the favored mechanisms for the UV upturn was
low-level star formation. Indeed, recent UV imaging surveys from the
{\it Galaxy Evolution Explorer} (\galex) have shown that a very
large fraction (at least 15\,\%) of early-type galaxies as
classified by the Sloan Digital Sky Survey (SDSS) exhibit signatures
of ongoing or recent star formation \citep{yetal05,setal07,ketal07}.
In Ultraviolet Imaging Telescope (UIT) images of the centre of M31,
however, \citet{betal85} had not detected point sources suggesting
that there is no evidence for the presence of main-sequence O stars
even though its UV colour is blue. Furthermore, \citet{ w82} claimed
that the absence of strong C\,{\small IV} and Si\,{\small IV} in the
FUV spectra of M31 indicates that  young stellar populations are not
present in the bulge of this galaxy.

What, then, is the origin of the UV upturn phenomenon? Early stellar
population models reproducing the red colours and spectral energy
distributions (SEDs) of early-type galaxies in the optical band
suggested that most stars in elliptical galaxies should be old and
metal-rich \citep[e.g.,][]{f72,t72}, but these models failed to
reproduce the strong observed UV flux. It is now generally
recognized that early-type galaxies also contain old stellar
components with temperatures exceeding 20,000 K. The leading
hypothesis suggests that the FUV flux originates from a minority
population in the helium-burning horizontal-branch (HB) phase and
its progeny, but there are two competing HB solutions: metal-poor
\citep*{ldz94,pl97,bg08} and metal-rich
\citep*{gr90,hdp92,bcf94,dor95,ydk97}. It is widely known that
metal-poor HB stars can be hot and become UV bright when they are
old. The oldest stars in this scenario are the most metal-poor and
are most likely to be found in galactic centres, where the UV flux
is strong, but the required age is older than that of typical Milky
Way globular clusters \citep{yi99}. On the other hand, metal-rich HB
stars \citep{bcf94,ydk97} generate UV sources by skipping the
asymptotic giant branch (AGB) phase and require a similar age as the
Milky Way halo. This phenomenon, dubbed the AGB manqu\'{e} stage
\citep{gr90}, is more pronounced at high values of helium abundance
\citep{hdp92,dor95}. Metal-rich stars thus become UV-bright sources.
A thorough review of these issues is available in \citet{o99}.

Based on recent observations, a number of studies have identified
peculiar globular clusters with extended horizontal branch (EHB)
stars attributable to the presence of super-helium-rich populations
\citep*[see e.g.][]{letal05, lgc07}. These studies also discussed
the discovery of numerous UV-bright globular clusters in the giant
elliptical galaxy M87 \citep[see e.g.][]{soklbbfr06}. Although the
origin of this helium enhancement is not yet fully understood, it
could be a source of the FUV flux in elliptical galaxies. Binary
stars offer another possibility: \citet{hanetal03}, who constructed
population synthesis models including binaries, concluded that most
UV sources in elliptical galaxies come from binary stars and that
there is no temporal evolution in FUV$-V$ colour \citep[see
e.g.][]{mhmn01,hpl07}. The other important candidate is post-AGB
(PAGB) stars \citep[e.g.][]{gr90,dor95,bfdd97,bbkf00}, which also
generate FUV radiation, but are thought to account for only 10 to
30~\% of the UV flux.

A crucial observational result is the dependence of the UV upturn on
the stellar population and dynamical properties of early-type
galaxies (\citealt{f83}; \citeauthor*{bbbfl88} \citeyear{bbbfl88},
hereafter \citeauthor{bbbfl88}). \citeauthor{bbbfl88} found
correlations of the UV$-$optical colours with both the Mg$_{2}$
index and the central velocity dispersion, based on 24 quiescent
early-type galaxies. However, a correlation opposite to the Burstein
relation has also been reported. \citet{retal05} constructed a
FUV$-r$ colour from \galex\ UV and SDSS optical imaging and reported
no correlation between the FUV$-r$ colour and the Mg$_{2}$ index,
even after attempting to exclude galaxies with an active galactic
nucleus (AGN) and star formation.

Almost all previous studies focused on integrated UV--line strength
relations. However, it is important to look at these relations as a
function of radius within galaxies, since this can provide a finer
discrimination between UV upturn theories. It has been known
that a feature of the UV upturn is the strongest in the centers of
early-type galaxies. This produces the positive UV colour gradient
that can be interpreted in terms of the difference in the stellar
populations \citep[see e.g.][]{o92, o99}. \citet{bfdd97} claimed
that the FUV light originates in a population with high metallicity
and enhanced helium. If the UV colour variation is indeed related
to the metal abundance, galaxies with larger metallicity gradients
should have larger UV colour gradients. \citet{oetal98} found no
correlation between the FUV$-B$ colour gradients and internal
metallicity gradients via the Mg$_2$ index, but they had only a
small sample.

To reconcile some of these discrepancies, we obtained
spatially-resolved galaxy data that can provide detailed information
on the stellar population distribution. By combining UV imaging
observations from \galex\ with \sauron\ integral-field early-type
galaxy spectroscopy, we re-investigate the relationships originally
suggested by \citeauthor{bbbfl88}. In a companion paper,
\citeauthor*{betal11} (\citeyear{betal11}, hereafter
\citeauthor{betal11}) used \galex\ and \sauron\ data to revisit the
integrated UV--line strength relations using identical apertures for
all quantities. They recovered correlations of the integrated
FUV$-V$ and FUV$-$NUV colours with the integrated \mgb\ line
strength index. Furthermore, they argued that most outliers are due
to galaxies exhibiting low-level star formation.

In this paper, we focus on the internal gradients of colour, {\it
measured} line strength and {\it derived} stellar population
property within each galaxy. In Section~\ref{sec:obs}, we present a
brief summary of the photometric and spectroscopic data and related
analyses. The UV--index relations are presented and discussed in
Sections~\ref{sec:results} and \ref{sec:discussion1}. In
Section~\ref{sec:discussion2}, we discuss the origin of the UV
upturn phenomenon based on several population synthetic models. We
summarize our results and discuss their implications in
Section~\ref{sec:conclusions}.

%
%
\section{OBSERVATIONS AND DATA REDUCTION}
\label{sec:obs}
\subsection{UV observations}
\label{sec:galex}
We observed 34 early-type galaxies from the {\tt SAURON} sample of
48 \citep{zetal02} with the medium-depth imaging mode of {\it GALEX}
in both FUV ($1350$--$1750$\,\AA) and NUV ($1750$--$2750$\,\AA), as
part of our own UV imaging survey of the \sauron\ sample ({\it
GALEX} guest investigator programmes GI1--109 and GI3--041) and the
{\it GALEX} Nearby Galaxy Survey \citep[NGS;][]{getal07}. The
typical exposure time per field was one orbit ($\approx 1700$\,s),
sufficient to exploit the two-dimensional nature of the images. Of
the 48 early-type galaxies in the \sauron\ sample, another 7
galaxies have only short exposures (typically 100--150\,s) in the
\galex\ All-sky Imaging Survey (AIS), too shallow for
spatially-resolved work. The remaining 7 galaxies have not been
observed with \galex, explaining the current sample of 34 as listed
in Table~\ref{tab:rgrad}. Details of the data reduction and analysis
for these 34 early-type galaxies are described more fully
\citeauthor*{jetal09} (\citeyear{jetal09}, hereafter
\citeauthor{jetal09}), so we only summarise our processing here.
Details of the {\it GALEX} instruments, pipeline and calibration are
described in \citet{maetal05} and \citet{moetal05, moetal07}.

The spatial resolution of the \galex\ images is approximately
$4\farcs5$ and $6\farcs0$ FWHM in FUV and NUV, respectively, sampled
with $1\farcs5\times1\farcs5$ pixels. Although the images were
delivered after pre-processing, we performed our own sky subtraction
by measuring the sky level in source-free regions of the images.
Additionally, we convolved the FUV data to the spatial resolution of
the NUV observations before any analysis, to avoid spurious colour
gradients in the central regions.  We also generated a mask
image to minimize contamination by nearby sources using SEXTRACTOR,
and carefully examined. The mask images are good in most cases, but
for a few cases, we had to mask some sources manually. In the case
of NGC\,4486, for example, we mask the northwestern quadrant to
minimize the effect of the jet, but we do not remove the likely
contamination by non-thermal UV emission. Finally, we carried out
surface photometry by measuring the surface brightness along
elliptical annuli using the ELLIPSE task within the STSDAS ISOPHOTE
package in IRAF (Image Reduction and Analysis Facility), after
masking out nearby sources.  For each image, the centre of the
isophotes was fixed to the centre of the light distribution and the
position angle (PA), ellipticity ($\epsilon$) and surface brightness
($\mu$) were fitted as a function of the radius. The ellipses were
fitted only to the NUV images, which have much better
signal-to-noise ratios (S/N) at all radii than the FUV images, and
these ellipses were then superimposed on the FUV images to derive
meaningful colours. We also de-reddened the colours for Galactic
extinction using values of {\it A}$\rm_{FUV}$~=~8.376~$\times$ ~{\it
E(B$-$V)} and {\it A}$\rm_{NUV}$~=~8.741~$\times$~{\it E(B$-$V)}
\citep{wetal05} and the reddening maps of \citet*{schetal98}.

\subsection{Optical observations}
\label{sec:mdm}

Ground-based optical imaging observations with the {\it Hubble Space
Telescope} ({\it HST}) filter $F555W$ (similar to Johnson $V$) were
obtained using the MDM Observatory 1.3-m McGraw-Hill Telescope over
several observing runs in 2002--2005, again as part of a larger
survey targeting the whole {\tt SAURON} galaxy sample. Details of
the MDM observations are described in \citeauthor*{fb11}
(\citeyear{fb11}, hereafter \citeauthor{fb11}) and the data were
reduced and calibrated in the standard manner. The field-of-view of
the MDM images is $17\farcm3\times17\farcm3$ with
$0\farcs508\times0\farcs508$ pixels, allowing for accurate sky
subtraction and proper sampling of the seeing. The seeing for the
observations was $\approx 1\farcs2$.

The MDM data were also convolved to the resolution of the {\it
GALEX} $NUV$ data for the current analysis. The surface brightness
profiles were measured after superimposing the NUV ellipses as for
the FUV data. The magnitudes at each isophote were then corrected
for Galactic extinction using a value of {\it R}$\rm_V$~=~3.1
\citep*{cetal89} and the reddening maps of \citet{schetal98}.

\subsection{SAURON linestrengths and kinematics}
\label{sec:sauron}

The \sauron\ observations were designed to determine the
two-dimensional stellar kinematics (\citeauthor*{eetal04}
\citeyear{eetal04}, hereafter \citeauthor{eetal04};
\citeauthor*{eetal07} \citeyear{eetal07}, hereafter
\citeauthor{eetal07}), stellar line strengths and populations
(\citeauthor*{ketal06} \citeyear{ketal06}, hereafter
\citeauthor{ketal06}; \citeauthor*{ketal10} \citeyear{ketal10},
hereafter \citeauthor{ketal10}) and ionised gas kinematics and
properties of 48 nearby early-type galaxies in the field and
clusters \citep {zetal02}, using the panoramic integral-field
spectrograph for the William Herschel Telescope (WHT) based on the
\sauron\ TIGER microlens concept . The observations and data
reduction are described in the indicated papers.

To explore the UV--line strength relations as a function of radius
within individual galaxies, we also imposed the ellipses from the
NUV images on the \sauron\ maps of {\it measured} line strengths
(\mgb, Fe5015 and \hb), {\it derived} (as opposed to measured)
stellar population properties ($\alpha$-enhancement [$\alpha$/Fe],
metallicity [Z/H] and age) and stellar velocity dispersion
($\sigma$). \citeauthor{ketal10} transformed the \mgb, Fe5015,
Fe5270 and \hb\ line strengths into stellar population properties
using the single stellar population (SSP) models of \citet{s07}.
This process is fully described in \citeauthor{ketal10}. It is
important to bear in mind that these parameters are derived from
only three spectral lines (sometimes four lines), owing to the
narrow bandwidth of the \sauron\ spectra, and that they are
``SSP-equivalent'' parameters, relying on the supposition that the
stars in each galaxy have formed in a single burst. This is an {\it
ad hoc} assumption; however, it allows comparisons between the
SSP-equivalent values of different sample galaxies. We use simple
luminosity-weighted line strength, population and velocity
dispersion averages for each isophote. The legitimacy of this method
is discussed in \citeauthor{betal11}.

%
%

\begin{table*}
\caption{Parameters of the best-fit linear UV--line strength (and
stellar property) relations for old galaxies only.}
\label{tab:linear_fits}
\begin{tabular}{@{}llrrrrrr}
\hline
Colour    & Index  & Slope & Unit & Zero-point & Unit  & Coefficient & Significance level \\
\hline
FUV$-V$   & \mgb             & $-0.48\pm0.03$  & \AA\ mag$^{-1}$   & $7.97\pm0.23$  & \AA  & -0.77 & $<$0.001 \\
          & Fe               & $-0.18\pm0.04$  & \AA\ mag$^{-1}$   & $6.33\pm0.29$  & \AA  & -0.43 & $<$0.001\\
          & \hb              & $ 0.13\pm0.01$  & \AA\ mag$^{-1}$   & $0.58\pm0.09$  & \AA  &  0.76 & $<$0.001\\
          & [$\alpha$/Fe]    & $-0.04\pm0.01$  &      mag$^{-1}$   & $0.55\pm0.05$  &     & -0.46 & $<$0.001\\
          & [Z/H]            & $-0.12\pm0.02$  &      mag$^{-1}$   & $0.79\pm0.08$  &      &  -0.59 & $<$0.001\\
          & age              & $-0.06\pm0.01$  &  Gyr mag$^{-1}$   & $1.56\pm0.05$  & Gyr  & -0.49 & $<$0.001\\
          & $\sigma_{\rm e}$ & $-0.14\pm0.02$  & km s$^{-1}$ mag$^{-1}$  & $3.15\pm0.09$  &  km s$^{-1}$  & -0.54 & $<$0.001\\
FUV$-$NUV & \mgb             & $-0.73\pm0.04$  & \AA\ mag$^{-1}$   & $5.06\pm0.05$  & \AA  & -0.86 & $<$0.001\\
          & Fe               & $-0.36\pm0.06$  & \AA\ mag$^{-1}$   & $5.09\pm0.07$  & \AA  & -0.59 & $<$0.001\\
          & \hb              & $ 0.19\pm0.02$  & \AA\ mag$^{-1}$   & $1.39\pm0.03$  & \AA  &  0.72 & $<$0.001\\
          & [$\alpha$/Fe]    & $-0.03\pm0.01$  &      mag$^{-1}$   & $0.33\pm0.01$  &      & -0.29 & 0.005\\
          & [Z/H]            & $-0.20\pm0.02$  &      mag$^{-1}$   & $0.18\pm0.02$  &      & -0.72 & $<$0.001\\
          & age              & $-0.08\pm0.01$  &  Gyr mag$^{-1}$   & $1.21\pm0.01$  & Gyr  & -0.50 & $<$0.001\\
          & $\sigma_{\rm e}$ & $-0.25\pm0.03$  & km s$^{-1}$ mag$^{-1}$  & $2.50\pm0.02$  &  km s$^{-1}$  & -0.68 & $<$0.001\\
\hline
\end{tabular}
\end{table*}

%
%
\begin{figure*}
\begin{center}
\includegraphics[width=0.80\textwidth,clip=]{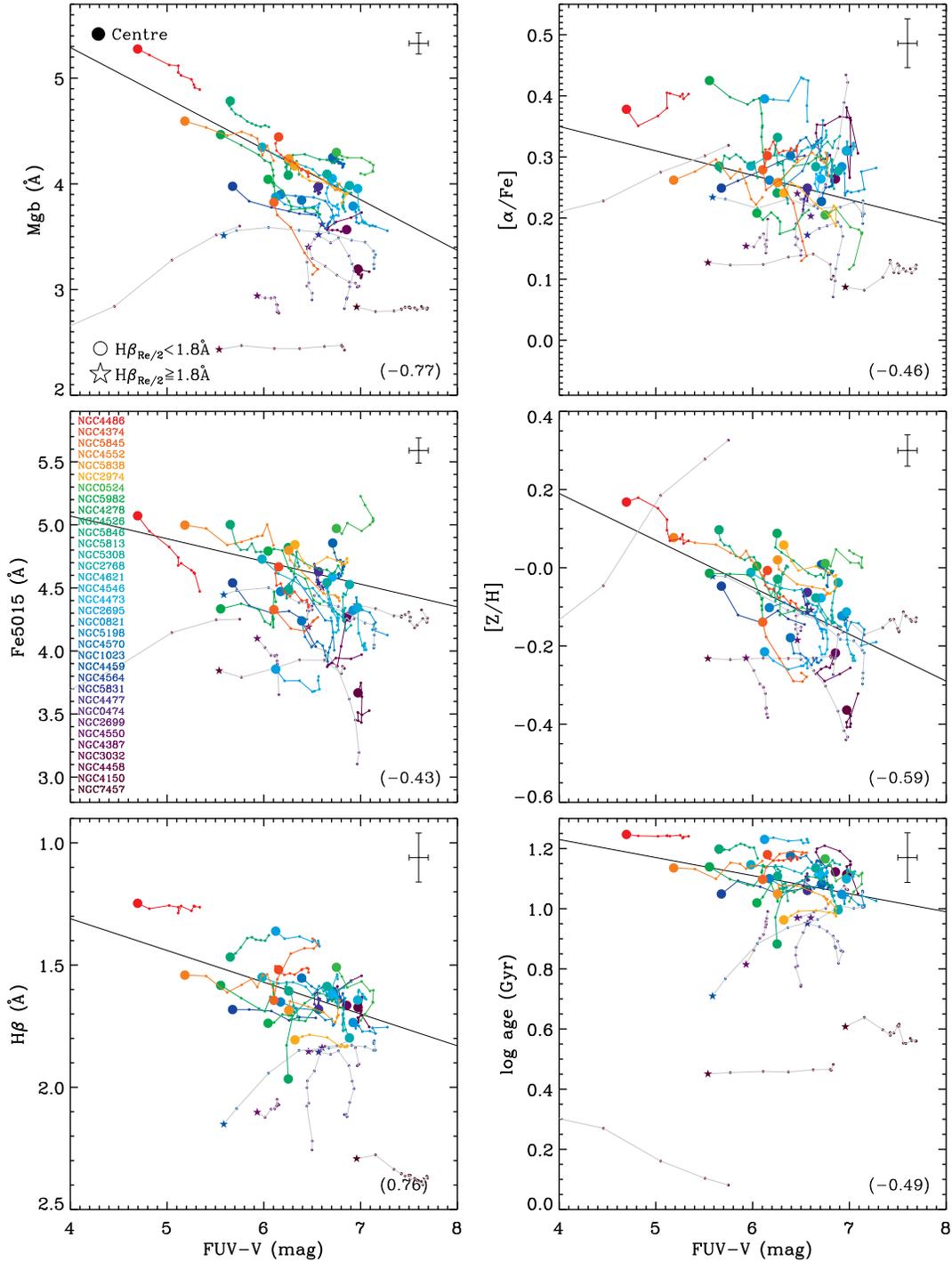}
\end{center}
\caption[]{UV upturn dependence on  {\it measured} line strengths
and {\it derived} stellar population properties. The \mgb\ ({\it
top-left}), Fe5015 ({\it middle-left}), \hb\ ({\it bottom-left}),
[$\alpha$/Fe] ({\it top-right}), [Z/H] ({\it middle-right}) and age
({\it bottom-right}) parameters are shown as a function of the
FUV$-V$ colour. We note that all  line strengths and {\it derived}
stellar population properties are based on the \sauron\ data
published in Paper~VI and Paper~XVII, respectively. The sample is
divided into quiescent early-type galaxies (circles, \hb$_{R_{\rm
e}/2}\,\le1.8$~\AA) and galaxies with recent star formation (stars,
\hb$_{R_{\rm e}/2}\,>1.8$~\AA). The solid line in each panel is a
linear fit to all of the quiescent galaxy profiles, weighting each
radius by its V-band luminosity, and the correlation coefficient is
reported in the bottom-right corner. The symbols are colour-coded
according to the integrated stellar velocity dispersion within one
effective radius ($\sigma_e$). All the points for an individual
galaxy are represented by a single colour, and the symbol for the
central point is larger.} \label{fig:fuvv}
\end{figure*}
%

%
%
\begin{figure*}
\begin{center}
\includegraphics[width=0.80\textwidth,clip=]{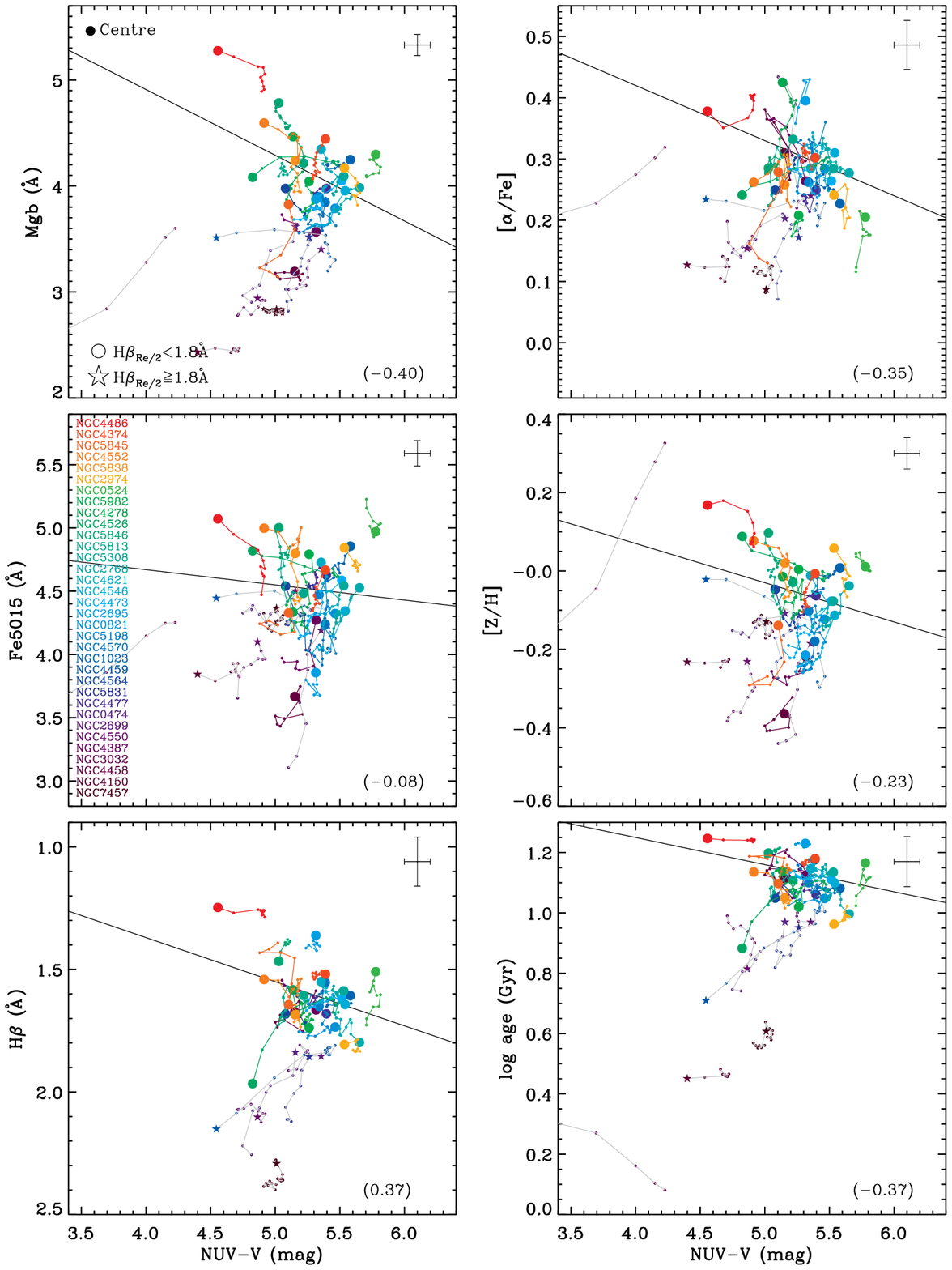}
\end{center}
\caption[]{Same as Figure~\ref{fig:fuvv} but for the NUV$-V$
colour.} \label{fig:nuvv}
\end{figure*}
%

%
%
\begin{figure*}
\begin{center}
\includegraphics[width=0.80\textwidth,clip=]{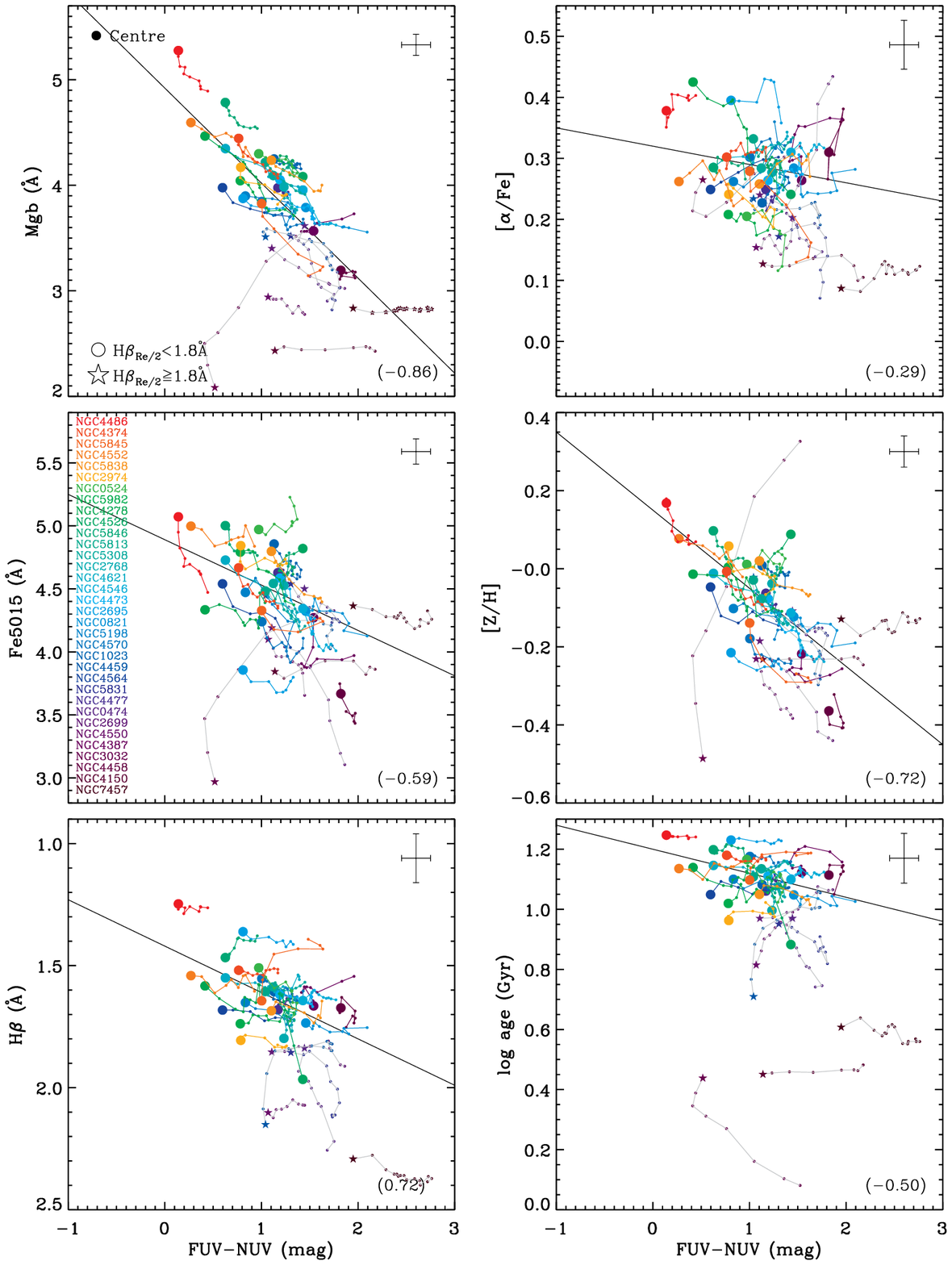}
\end{center}
\caption[]{Same as Figure~\ref{fig:fuvv} but for the FUV$-$NUV
colour.} \label{fig:fnuv}
\end{figure*}
%

%
%
\section{UV RELATIONS}
\label{sec:results}
As argued in the introduction, spatially-resolved galaxy data
provide valuable information for discriminating between UV upturn
theories. The internal stellar population gradients within each
galaxy can yield their own ``local'' relations, rather than relying
on UV--stellar population correlations from the integrated
measurements of many galaxies (see e.g. \citeauthor{betal11}). When
exploring correlations between stellar population parameters and
multi-band photometry, our main advantage is that all photometric
and spectroscopic measurements are derived using the same ellipses.
We note that all  stellar absorption line strengths (\mgb, Fe5015
and \hb),  stellar population properties ([$\alpha$/Fe], [Z/H] and
age) and stellar kinematics ($\sigma$) are based on the \sauron\
data published in \citeauthor{ketal06}, \citeauthor{ketal10},
\citeauthor{eetal04} and \citeauthor{eetal07}.

Strong Balmer absorption lines betray the presence of young stars.
Based on the prediction of various stellar population synthesis
models  \citep*[see e.g.][]{tmb03,s07}, an \hb\ line strength of
$\approx 1.8$\,\AA\ can be used to divide the sample into galaxies
with effectively only old stellar populations and those which have a
contribution from younger stars. To minimise the effects of galaxies
with recent star formation on the correlations, we represent
galaxies with an integrated \hb\ line strength within half an
effective radius \hb$_{R_{\rm e}/2}\,>1.8$~\AA\ (see Table~1 in
\citeauthor{betal11}) as star symbols (Figures 1--7, 10 and 11). The
symbols in Figures 1--7, 10--12 are also colour-coded according to
the integrated stellar velocity dispersion within one effective
radius ($\sigma_{\rm e}$; Table~A1 in \citeauthor{ketal10}),
descending from red to purple.

\subsection{Radial profiles}
\label{sec:radial}

The left-hand panels of Figure~\ref{fig:fuvv} show the correlations
between the FUV$-V$ colour and the  line strength indices \mgb,
Fe5015 and \hb\ as a function of radius. Linear fits to all the
quiescent galaxy (\hb$_{R_{\rm e}/2}\,\le1.8$~\AA) profiles, after
weighting each radius by its luminosity at $V$ band, are shown as
solid lines. The value of the correlation coefficient is reported in
the bottom-right corner of each panel, and the parameters of the
best-fit lines are reported in Table~\ref{tab:linear_fits}, with
errors properly taken into account.

The top-left panel of Figure~\ref{fig:fuvv} shows the
(FUV$-V$)--\mgb\ ``Burstein'' relation. This relation shows the
tightest correlation among the different line strength indices, as
discussed in many previous studies. The slope of a linear fit to the
integrated properties of the quiescent galaxies (including the 7 AIS
galaxies; see Table~1 of \citeauthor{betal11}) is -0.40$\pm$0.09
\AA\ mag$^{-1}$, whereas the slope measured here based on the
luminosity-weighted galaxy profiles (solid line; global gradient)
has a value of -0.48$\pm$0.03 \AA\ mag$^{-1}$. These two slopes are
thus consistent within the errors and our results are in good
agreement with those of \citeauthor{betal11}. All of the quiescent
galaxies exhibit bluer FUV$-V$ colours at smaller radii, suggesting
that the FUV light is more centrally-concentrated than the optical
light, also in agreement with earlier studies \citep[see
e.g.][]{oetal98,o99}.

We stress that galaxies with young stellar populations (\hb$_{R_{\rm
e}/2}>1.8$\,\AA, star symbols) show different trends (either
opposite trends or constant \mgb\ values as FUV$-V$ colours are
varied) than quiescent galaxies, so the relation becomes less
distinct. These galaxies also have lower \mgb\ values. It is clear
therefore that a tight (FUV$-V$)--\mgb\ relation is only present for
galaxies which show no sign of on-going or recent star formation.

The middle-left panel of Figure~\ref{fig:fuvv} shows the relation of
FUV$-V$ with Fe5015. This relation is much weaker, as found by
\citet{o99} and \citeauthor{betal11}. This implies that the UV
strength is more strongly linked to the behavior of the lighter
elements (e.g. Mg), not the iron peak. \footnote{Admittedly, our
analysis is limited to Fe5015 and Mgb as iron-peak and  magnesium
strength indicators due to the narrow wavelength coverage of the
SAURON instrument. A wider wavelength coverage is desired to
constrain them more robustly.} We also note that galaxies with a
young population appear at lower Fe values.

In the case of \hb\ (bottom-left panel of Figure~\ref{fig:fuvv}),
there is a clear correlation with the FUV$-V$ colour. A tight
correlation is, again, only present among quiescent galaxies. The
separation of young galaxies (stars) from old ones (circles) is
however somewhat artificial in this plot, as we use \hb\ itself as
the dividing criterion. The observed variation of \hb\ with FUV$-V$
is also expected from the (FUV$-V$)--\mgb\ correlation. According to
stellar population models, differences in metallicity cause
variations in the \hb\ line strength. For example, in the models of
\citet{tmb03}, an \mgb\ value of 5.16 \AA\ and an age of 12~Gyr
correspond to an \hb\ line strength of 1.50 \AA, whereas for \mgb
\,$=$\,3.11 \AA\ and age\,=\,12~Gyr the \hb\ line strength increases
to 1.77 \AA. The observed trend can thus easily be explained by the
effect of metallicity on \hb.

The right-hand panels of Figure~\ref{fig:fuvv} show the correlations
between the FUV$-V$ colour and the  stellar population properties
([$\alpha$/Fe], [Z/H] and age) as a function of radius. The
parameters of the best-fit lines and the values of the correlation
coefficients are again listed in Table~\ref{tab:linear_fits}. Only
[Z/H] shows a modest correlation with FUV. It is surprising that
[$\alpha$/Fe] does not show a clear correlation, considering the
tight correlation of \mgb\ with FUV.

Analogously to Figure~\ref{fig:fuvv}, Figure~\ref{fig:nuvv} shows
the correlations between the NUV$-V$ colour and the line strength
indices (\mgb, Fe5015 and \hb) and stellar population parameters
([$\alpha$/Fe], [Z/H] and age) as a function of radius. In contrast
to Figure~\ref{fig:fuvv}, we note no correlation between the NUV$-V$
colour and those quantities. The UV upturn phenomenon clearly
dominates the spectral energy distribution only at wavelengths
blueward of 2000\,\AA. We therefore believe that these trends are
caused by two main effects: young stars and line blanketing. Recent
star formation ($t\la1$~Gyr) has a greater effect in the NUV than in
the FUV, as only on-going star formation ($t\la0.1$~Gyr) can
contribute to the FUV light. On the other hand, line blanketing is
significant in the NUV as metallicity increases, as a large number
of partially overlapping absorption lines can suppress the
continuum. These are probably the main reasons why the FUV
correlations become weaker in the NUV. Nevertheless, galaxies with
extremely high \mgb\ line strength and a strong UV upturn also show
bluer NUV$-V$ colours.

%
%
\begin{figure*}
\begin{center}
\includegraphics[width=0.80\textwidth,clip=]{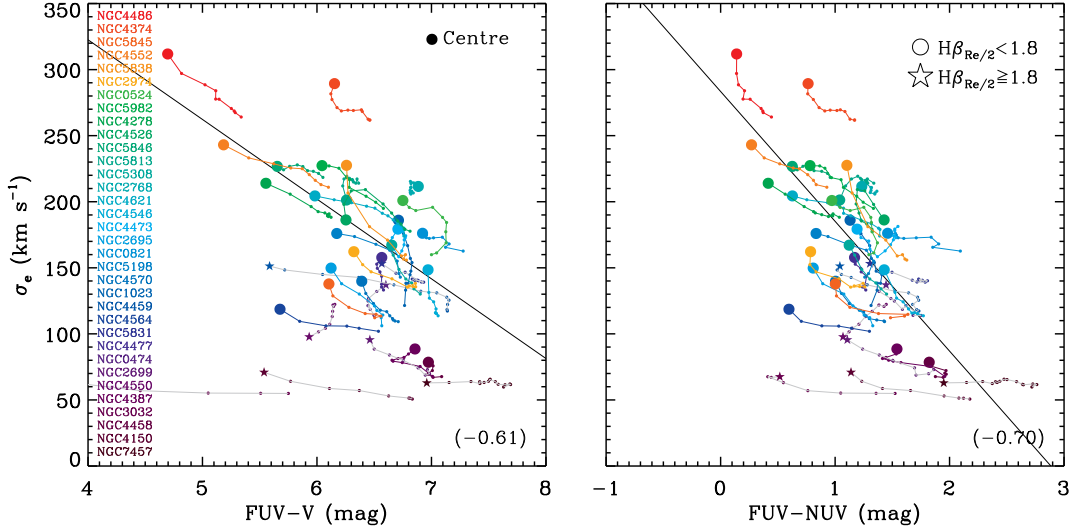}
\end{center}
\caption[]{Same as Figure~\ref{fig:fuvv} but for the FUV$-$NUV and
FUV$-V$ colours as a function of the integrated stellar velocity
dispersion within one effective radius $\sigma_{\rm e}$.}
\label{fig:vd}
\end{figure*}

Figure~\ref{fig:fnuv} shows the correlations between the FUV$-$NUV
colour and the line strength indices (\mgb, Fe5015 and \hb) and
stellar population parameters ([$\alpha$/Fe], [Z/H] and age) as a
function of radius. The luminosity-weighted linear fits to the data
with \hb$_{R_{\rm e}/2}\,\le1.8$~\AA\ are shown as solid lines and
the parameters of the best-fit lines are again reported in
Table~\ref{tab:linear_fits}. The value of the correlation
coefficient is also reported in the bottom-right corner of each
panel and Table~\ref{tab:linear_fits}. Note that the UV--line
strength correlations are much more pronounced when the FUV$-V$
colour is replaced by the FUV$-$NUV colour (see also
\citealt{detal07} and \citeauthor{betal11}). For example, the
(FUV$-$NUV)--\mgb\ correlation is tighter than that of
(FUV$-V$)--\mgb. We mentioned above that line blanketing is
substantial in the NUV and increases with metallicity. This is
likely the reason why FUV$-$NUV correlations are much tighter than
those with FUV$-V$. Another advantage of using FUV$-$NUV rather than
FUV$-V$ is that both FUV and NUV magnitudes are measured with the
same instrument, and uncertainties in the UV to optical flux
calibration are avoided.

\subsection{Stellar velocity dispersion}
\label{sec:vd}
It is well-known that there is a tight correlation between the
stellar velocity dispersion ($\sigma$) and the Mg index \citep*[see
e.g.][]{tdfb81,bbf93,cbdmsw99,betal03}. This is a well-studied
relation linking a dynamical quantity $\sigma$ with a quantity (the
Mg index) that depends on the stellar populations. A number of
studies have implied a correlation of the UV-upturn phenomenon with
the stellar velocity dispersion \citep[see e.g.][]{bbbfl88,detal07},
a good measure of a galaxy's gravitational potential depth and total
dynamical mass.

Figure~\ref{fig:vd} shows the colour--$\sigma_{\rm e}$ relations.
Comparing to Figures~\ref{fig:fuvv} and \ref{fig:fnuv} (top-left),
one can see that the (FUV$-V$)--\mgb\ and (FUV$-$NUV)--\mgb\
relations are tighter than the (FUV$-V$)--$\sigma_{\rm e}$ and
(FUV$-$NUV)--$\sigma_{\rm e}$ relations, confirming the results of
\citeauthor{bbbfl88}.

%
%
\begin{figure}
\begin{center}
\includegraphics[width=0.45\textwidth,clip=]{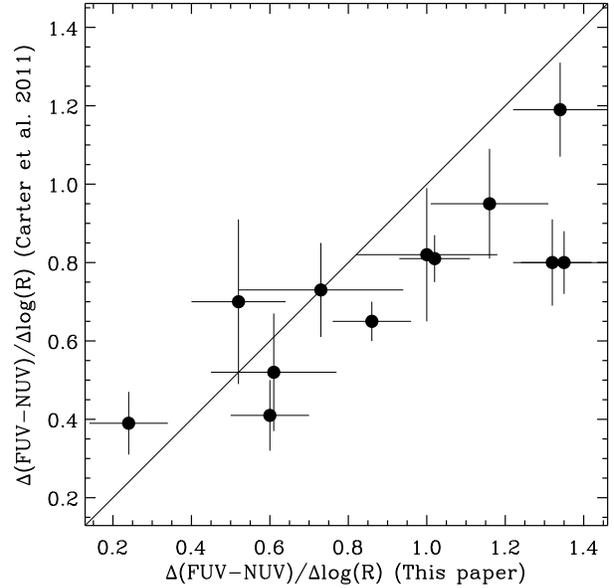}
\end{center}
\caption[]{Comparison of our (FUV$-$NUV) radial colour gradients
with those of \citet{cetal11}.} \label{fig:comp}
\end{figure}
%

%
%
\begin{table*}
\caption{Logarithmic internal radial gradients.} \label{tab:rgrad}
\begin{tabular}{@{}lrrrrrrrr}
\hline
Galaxy & $\frac{\Delta \rm FUV-V}{\Delta \rm log(R)}$ &
$\frac{\Delta \rm FUV-NUV}{\Delta \rm log(R)}$ &
         $\frac{\Delta \rm Mgb}{\Delta \rm log(R)}$   & $\frac{\Delta \rm Fe5015}{\Delta \rm log(R)}$  &
         $\frac{\Delta \rm Hb}{\Delta \rm log(R)}$   & $\frac{\Delta \rm [\alpha/Fe]}{\Delta \rm log(R)}$ &
         $\frac{\Delta \rm [Z/H]}{\Delta \rm log(R)}$  & $\frac{\Delta \rm log(age)}{\Delta \rm log(R)}$ \\
 & (mag dex$^{-1}$) & (mag dex$^{-1}$) & (\AA\ dex$^{-1}$) &  (\AA\ dex$^{-1}$) &
     (\AA\ dex$^{-1}$) & &  & (Gyr dex$^{-1}$) \\
(1) & (2) & (3) & (4) & (5) & (6) & (7) & (8) & (9)\\
\hline
%
NGC0474 &   -0.18   $\pm$   0.16    &   0.73    $\pm$   0.21    &   -1.70   $\pm$   0.26    &   -0.30   $\pm$   0.27    &   0.89    $\pm$   0.26    &   -0.17   $\pm$   0.11    &   -0.14   $\pm$   0.11    &   -0.52   $\pm$   0.14    \\
NGC0524 &   0.66    $\pm$   0.14    &   0.73    $\pm$   0.23    &   -0.44   $\pm$   0.21    &   0.44    $\pm$   0.22    &   0.38    $\pm$   0.21    &   -0.18   $\pm$   0.08    &   0.04    $\pm$   0.08    &   -0.21   $\pm$   0.09    \\
NGC0821 &   0.44    $\pm$   0.16    &   1.01    $\pm$   0.26    &   -0.40   $\pm$   0.21    &   -0.38   $\pm$   0.21    &   0.05    $\pm$   0.21    &   -0.02   $\pm$   0.10    &   -0.12   $\pm$   0.10    &   -0.02   $\pm$   0.10    \\
NGC1023 &   0.14    $\pm$   0.05    &   0.47    $\pm$   0.10    &   -0.06   $\pm$   0.16    &   -0.45   $\pm$   0.16    &   0.03    $\pm$   0.16    &   0.08    $\pm$   0.07    &   -0.07   $\pm$   0.07    &   0.02    $\pm$   0.08    \\
NGC2695 &   1.06    $\pm$   0.13    &   1.24    $\pm$   0.17    &   -0.26   $\pm$   0.26    &   -0.19   $\pm$   0.28    &   0.11    $\pm$   0.26    &   -0.02   $\pm$   0.10    &   -0.05   $\pm$   0.10    &   -0.01   $\pm$   0.05    \\
NGC2699 &   1.33    $\pm$   0.18    &   1.81    $\pm$   0.23    &   -0.89   $\pm$   0.26    &   -2.50   $\pm$   0.29    &   0.10    $\pm$   0.26    &   0.57    $\pm$   0.11    &   -0.63   $\pm$   0.11    &   0.28    $\pm$   0.14    \\
NGC2768 &   -0.06   $\pm$   0.06    &   0.24    $\pm$   0.10    &   -0.43   $\pm$   0.17    &   -0.41   $\pm$   0.17    &   -0.18   $\pm$   0.17    &   -0.03   $\pm$   0.08    &   -0.17   $\pm$   0.08    &   0.07    $\pm$   0.09    \\
NGC2974 &   1.31    $\pm$   0.09    &   1.19    $\pm$   0.16    &   -0.65   $\pm$   0.23    &   -0.19   $\pm$   0.23    &   0.09    $\pm$   0.22    &   -0.12   $\pm$   0.09    &   -0.21   $\pm$   0.09    &   0.08    $\pm$   0.11    \\
NGC3032 &   7.59    $\pm$   0.11    &   2.67    $\pm$   0.15    &   3.62    $\pm$   0.27    &   3.18    $\pm$   0.29    &   1.52    $\pm$   0.26    &   0.22    $\pm$   0.11    &   2.01    $\pm$   0.12    &   -0.90   $\pm$   0.08    \\
NGC4150 &   3.29    $\pm$   0.15    &   2.74    $\pm$   0.20    &   0.01    $\pm$   0.26    &   0.10    $\pm$   0.26    &   0.11    $\pm$   0.26    &   -0.05   $\pm$   0.10    &   -0.01   $\pm$   0.10    &   0.05    $\pm$   0.09    \\
NGC4278 &   0.87    $\pm$   0.05    &   1.02    $\pm$   0.09    &   -1.01   $\pm$   0.18    &   -0.01   $\pm$   0.19    &   -0.02   $\pm$   0.18    &   -0.23   $\pm$   0.08    &   -0.26   $\pm$   0.08    &   0.08    $\pm$   0.08    \\
NGC4374 &   0.75    $\pm$   0.05    &   0.86    $\pm$   0.10    &   -0.64   $\pm$   0.17    &   -0.53   $\pm$   0.17    &   0.01    $\pm$   0.17    &   0.02    $\pm$   0.07    &   -0.17   $\pm$   0.07    &   -0.02   $\pm$   0.07    \\
NGC4387 &   -0.10   $\pm$   0.14    &   0.43    $\pm$   0.18    &   0.31    $\pm$   0.23    &   -0.48   $\pm$   0.24    &   -0.20   $\pm$   0.22    &   0.22    $\pm$   0.10    &   -0.08   $\pm$   0.10    &   0.14    $\pm$   0.10    \\
NGC4458 &   0.09    $\pm$   0.16    &   0.36    $\pm$   0.21    &   0.04    $\pm$   0.26    &   -0.53   $\pm$   0.27    &   0.17    $\pm$   0.26    &   0.12    $\pm$   0.13    &   0.06    $\pm$   0.13    &   -0.07   $\pm$   0.13    \\
NGC4459 &   2.82    $\pm$   0.13    &   1.58    $\pm$   0.15    &   -0.63   $\pm$   0.17    &   -0.84   $\pm$   0.17    &   -0.53   $\pm$   0.17    &   -0.05   $\pm$   0.07    &   -0.51   $\pm$   0.07    &   0.56    $\pm$   0.09    \\
NGC4473 &   0.08    $\pm$   0.12    &   0.61    $\pm$   0.16    &   -0.58   $\pm$   0.16    &   -0.70   $\pm$   0.16    &   0.01    $\pm$   0.16    &   0.05    $\pm$   0.07    &   -0.22   $\pm$   0.07    &   0.02    $\pm$   0.08    \\
NGC4477 &   0.10    $\pm$   0.08    &   0.24    $\pm$   0.13    &   -0.16   $\pm$   0.26    &   -0.17   $\pm$   0.26    &   -0.06   $\pm$   0.26    &   0.00    $\pm$   0.11    &   -0.10   $\pm$   0.11    &   0.10    $\pm$   0.12    \\
NGC4486 &   1.22    $\pm$   0.13    &   0.85    $\pm$   0.21    &   -0.87   $\pm$   0.24    &   -1.15   $\pm$   0.24    &   0.02    $\pm$   0.23    &   0.10    $\pm$   0.09    &   -0.28   $\pm$   0.09    &   -0.01   $\pm$   0.02    \\
NGC4526 &   1.25    $\pm$   0.12    &   0.07    $\pm$   0.13    &   0.16    $\pm$   0.21    &   -0.40   $\pm$   0.21    &   -0.49   $\pm$   0.21    &   0.11    $\pm$   0.08    &   -0.20   $\pm$   0.08    &   0.34    $\pm$   0.11    \\
NGC4546 &   0.20    $\pm$   0.07    &   0.67    $\pm$   0.11    &   -0.63   $\pm$   0.22    &   -0.56   $\pm$   0.23    &   0.03    $\pm$   0.22    &   0.01    $\pm$   0.11    &   -0.22   $\pm$   0.11    &   0.01    $\pm$   0.11    \\
NGC4550 &   0.42    $\pm$   0.12    &   0.86    $\pm$   0.16    &   -0.36   $\pm$   0.23    &   -0.66   $\pm$   0.24    &   -0.11   $\pm$   0.23    &   0.04    $\pm$   0.11    &   -0.32   $\pm$   0.11    &   0.34    $\pm$   0.13    \\
NGC4552 &   1.91    $\pm$   0.04    &   1.35    $\pm$   0.11    &   -0.68   $\pm$   0.23    &   -0.13   $\pm$   0.24    &   0.04    $\pm$   0.22    &   -0.11   $\pm$   0.09    &   -0.18   $\pm$   0.09    &   0.05    $\pm$   0.08    \\
NGC4564 &   2.49    $\pm$   0.20    &   2.05    $\pm$   0.23    &   -0.75   $\pm$   0.35    &   -1.10   $\pm$   0.37    &   0.12    $\pm$   0.34    &   0.16    $\pm$   0.15    &   -0.22   $\pm$   0.15    &   -0.01   $\pm$   0.17    \\
NGC4570 &   0.64    $\pm$   0.08    &   0.87    $\pm$   0.10    &   -0.38   $\pm$   0.21    &   -0.33   $\pm$   0.22    &   0.16    $\pm$   0.21    &   0.04    $\pm$   0.10    &   -0.09   $\pm$   0.10    &   -0.06   $\pm$   0.09    \\
NGC4621 &   1.32    $\pm$   0.06    &   1.32    $\pm$   0.10    &   -0.67   $\pm$   0.17    &   -0.86   $\pm$   0.17    &   -0.02   $\pm$   0.17    &   0.09    $\pm$   0.08    &   -0.23   $\pm$   0.08    &   0.04    $\pm$   0.07    \\
NGC5198 &   1.40    $\pm$   0.18    &   1.22    $\pm$   0.24    &   -0.68   $\pm$   0.29    &   -1.28   $\pm$   0.30    &   0.25    $\pm$   0.29    &   0.25    $\pm$   0.13    &   -0.22   $\pm$   0.13    &   -0.02   $\pm$   0.14    \\
NGC5308 &   0.67    $\pm$   0.13    &   0.86    $\pm$   0.17    &   -0.50   $\pm$   0.29    &   -0.95   $\pm$   0.30    &   -0.03   $\pm$   0.29    &   0.11    $\pm$   0.13    &   -0.20   $\pm$   0.13    &   0.06    $\pm$   0.14    \\
NGC5813 &   0.22    $\pm$   0.06    &   0.52    $\pm$   0.12    &   -0.02   $\pm$   0.19    &   0.54    $\pm$   0.19    &   0.27    $\pm$   0.19    &   -0.18   $\pm$   0.08    &   0.13    $\pm$   0.08    &   -0.17   $\pm$   0.08    \\
NGC5831 &   0.58    $\pm$   0.13    &   1.00    $\pm$   0.18    &   -1.45   $\pm$   0.21    &   -0.50   $\pm$   0.22    &   0.62    $\pm$   0.21    &   -0.20   $\pm$   0.09    &   -0.26   $\pm$   0.09    &   -0.16   $\pm$   0.11    \\
NGC5838 &   1.29    $\pm$   0.27    &   1.32    $\pm$   0.38    &   -0.69   $\pm$   0.26    &   -0.95   $\pm$   0.27    &   0.05    $\pm$   0.26    &   0.07    $\pm$   0.11    &   -0.20   $\pm$   0.11    &   -0.11   $\pm$   0.13    \\
NGC5845 &   1.34    $\pm$   0.33    &   1.82    $\pm$   0.42    &   -1.88   $\pm$   0.31    &   -0.21   $\pm$   0.31    &   -0.60   $\pm$   0.29    &   -0.41   $\pm$   0.13    &   -0.37   $\pm$   0.13    &   0.14    $\pm$   0.10    \\
NGC5846 &   0.79    $\pm$   0.04    &   0.60    $\pm$   0.10    &   -0.49   $\pm$   0.18    &   -0.48   $\pm$   0.19    &   -0.19   $\pm$   0.18    &   0.00    $\pm$   0.08    &   -0.17   $\pm$   0.08    &   0.01    $\pm$   0.06    \\
NGC5982 &   1.11    $\pm$   0.13    &   1.16    $\pm$   0.15    &   -0.72   $\pm$   0.23    &   -0.90   $\pm$   0.23    &   -0.08   $\pm$   0.22    &   -0.02   $\pm$   0.09    &   -0.25   $\pm$   0.09    &   0.08    $\pm$   0.11    \\
NGC7457 &   1.00    $\pm$   0.13    &   1.09    $\pm$   0.15    &   0.03    $\pm$   0.14    &   -0.15   $\pm$   0.14    &   0.15    $\pm$   0.14    &   0.04    $\pm$   0.05    &   0.03    $\pm$   0.05    &   -0.09   $\pm$   0.06    \\
\hline
\end{tabular}
Columns: (1) Galaxy identifier. (2)--(3): Radial colour gradients.
(4)--(9): Radial line strength and stellar population gradients.
\end{table*}

%
%
\begin{figure*}
\begin{center}
\includegraphics[width=0.85\textwidth,clip=m]{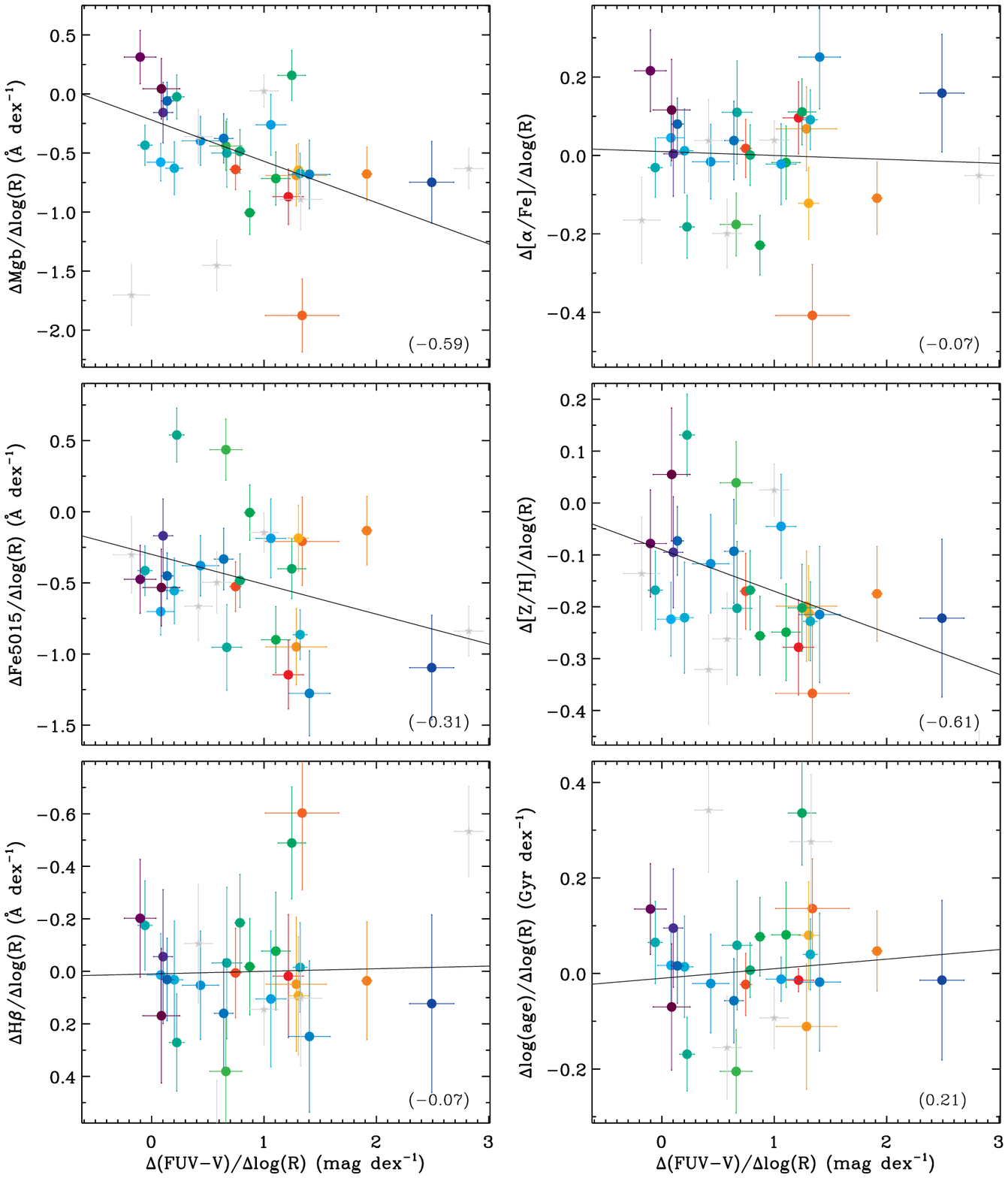}
\end{center}
\caption[]{Same as Figure~\ref{fig:fuvv} but for the FUV$-V$
logarithmic internal radial colour gradient versus the {\it
measured} and {\it derived} stellar population gradients of the
sample galaxies. The gradients of \mgb\ ({\it top-left}), Fe5015
({\it middle-left}), \hb\ ({\it bottom-left}), [$\alpha$/Fe] ({\it
top-right}), [Z/H] ({\it middle-right}) and age ({\it bottom-right})
are shown as a function of the FUV$-V$ colour gradient. Quiescent
galaxies (\hb$_{R_{\rm e}/2}\,\le1.8$~\AA) are shown as solid
circles, while galaxies with a young population are shown as grey
stars. The solid line in each panel is a fit to the quiescent
galaxies, and the correlation coefficient is reported in the
bottom-right corner.} \label{fig:fuvv_grad}
\end{figure*}
\subsection{Internal gradients}
\label{sec:gradient}

From the radial surface brightness,  line strength and stellar
population profiles, we derive logarithmic internal radial gradients
for each galaxy using least-square fits. The radial range of the
fits generally extends from the FWHM of the UV point spread function
to half the effective radius ($R_{\rm e}/2$). The radial gradients
are thus defined by their global slopes. In addition, to estimate
the gradient uncertainties, we performed Monte-Carlo simulations
comprising 10$^{3}$ realisations, adding noise compatible with the
observations to each data point.  In Figure~\ref{fig:comp}, we
show the comparison of our (FUV$-$NUV) radial colour gradients with
those of \citet{cetal11}. The agreement is generally good: the mean
of the absolute value of the (FUV$-$NUV) radial colour gradient
difference is $\approx$0.55 mag\,dex$^{-1}$. The largest difference
is found for NGC\,4552 likely due to the determination of the radial
range of the fits. We adopted the same radial range to derive the
internal radial gradients for each galaxy. The logarithmic internal
radial gradients of the UV colours, line strengths and stellar
population properties are listed in Table~\ref{tab:rgrad}.

We plot the internal gradients of the  line strength indices and the
stellar population parameters as a function of the FUV$-V$ internal
colour gradients in Figure~\ref{fig:fuvv_grad}. To prevent any
dilution of the correlation attributable to galaxies with recent
star formation, we exclude galaxies with evidence of young stars
(\hb$_{R_{\rm e}/2}\,>1.8$~\AA) from the fit and plot them as grey
stars. The value of the correlation coefficient is reported in the
bottom-right corner of each panel, and the parameters of the
best-fit lines are reported in Table~\ref{tab:linear_fits2}.

%
%

\begin{table*}
\caption{Parameters of the best-fit linear radial gradients for old
galaxies only.} \label{tab:linear_fits2}
\begin{tabular}{@{}llrrrrrr}
\hline
Colour    & Index  & Slope & Zero-point  & Coefficient  & Significance level\\
\hline
FUV$-V$/$\Delta$log(R)  & \mgb/$\Delta$log(R)           & $-0.35\pm0.13$  & $-0.22\pm0.12$  & -0.59 &  0.001\\
                        & Fe/$\Delta$log(R)             & $-0.21\pm0.16$  & $-0.30\pm0.15$  & -0.31 & 0.14\\
                        & \hb /$\Delta$log(R)           & $-0.01\pm0.09$  & $ 0.01\pm0.07$  & -0.07 & 0.67\\
                        & [$\alpha$/Fe]/$\Delta$log(R)  & $-0.01\pm0.05$  & $ 0.01\pm0.05$  & -0.07 & 0.68\\
                        & [Z/H] /$\Delta$log(R)         & $-0.08\pm0.04$  & $-0.09\pm0.04$  &  0.61 & $<$0.001\\
                        & log age/$\Delta$log(R)        & $ 0.02\pm0.04$  & $-0.01\pm0.04$  &  0.21 & 0.28\\
                    & log$\sigma_{\rm e}$/$\Delta$log(R)& $-0.30\pm0.04$  & $-0.12\pm0.04$  & -0.21& 0.27\\
FUV$-$NUV/$\Delta$log(R)  & \mgb/$\Delta$log(R)         & $-0.71\pm0.16$  & $ 0.10\pm0.14$  & -0.83 & $<$0.001\\
                          & Fe/$\Delta$log(R)           & $-0.31\pm0.25$  & $-0.20\pm0.23$  & -0.31 & 0.14\\
                          & \hb /$\Delta$log(R)         & $ 0.13\pm0.13$  & $-0.10\pm0.11$  &  0.44 & 0.018\\
                          & [$\alpha$/Fe]/$\Delta$log(R)& $-0.07\pm0.08$  & $ 0.05\pm0.07$  & -0.28 & 0.18\\
                          & [Z/H] /$\Delta$log(R)       & $-0.11\pm0.06$  & $-0.06\pm0.06$  &  0.54 & 0.004\\
                          & log age/$\Delta$log(R)      & $-0.03\pm0.06$  & $ 0.04\pm0.06$  & -0.20 & 0.31\\
                    & log$\sigma_{\rm e}$/$\Delta$log(R)& $-0.08\pm0.06$  & $-0.09\pm0.05$  & -0.30 & 0.15\\
\hline
\end{tabular}
\end{table*}

%
%
\begin{figure*}
\begin{center}
\includegraphics[width=0.85\textwidth,clip=m]{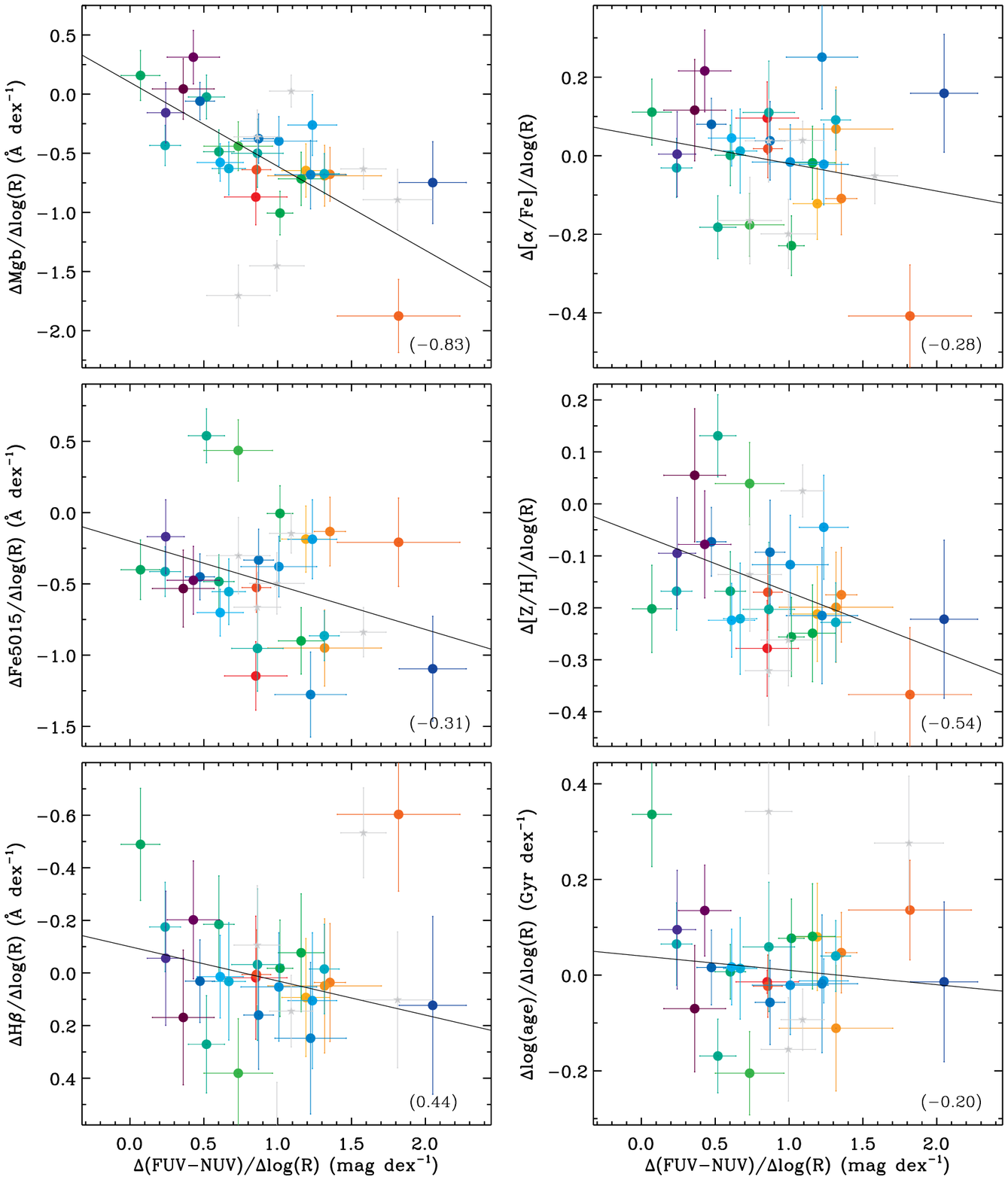}
\end{center}
\caption[]{Same as Figure~\ref{fig:fuvv_grad} but for the FUV$-$NUV
colour.} \label{fig:fnuv_grad}
\end{figure*}

%
%
\begin{figure*}
\begin{center}
\includegraphics[width=0.85\textwidth,clip=m]{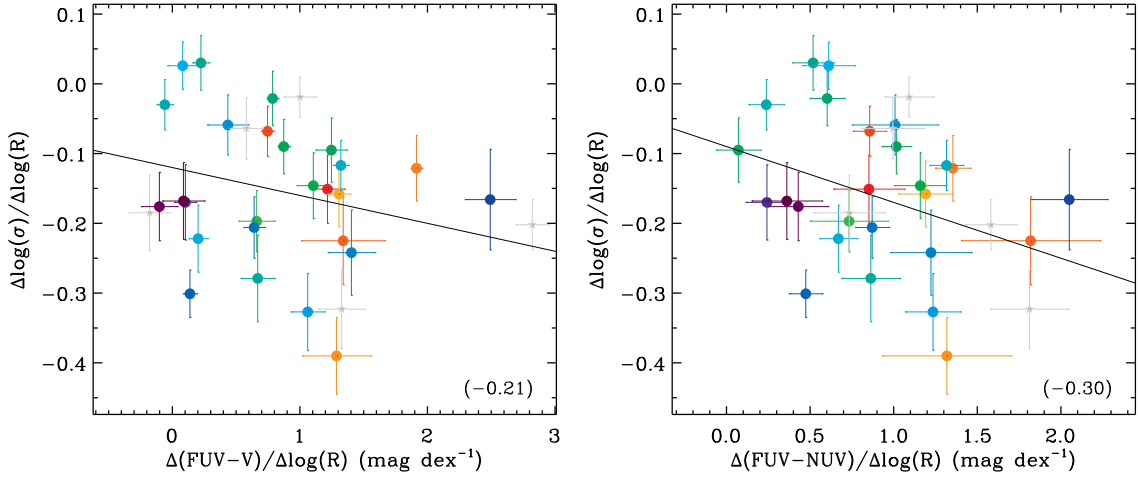}
\end{center}
\caption[]{Same as Figures~\ref{fig:fuvv_grad} and
\ref{fig:fnuv_grad} but for the stellar velocity dispersion.}
\label{fig:vd_grad}
\end{figure*}

%
%
\begin{table*}
\caption{Internal gradients.} \label{tab:grad}
\begin{tabular}{@{}lrrrrrrr}
\hline Galaxy & $\frac{\Delta \rm log\,\sigma}{\Delta \rm FUV-V}$ &
$\frac{\Delta \rm log\,\sigma}{\Delta \rm FUV-NUV}$ &
         $\frac{\Delta \rm Mgb}{\Delta \rm FUV-V}$ & $\frac{\Delta \rm Mgb}{\Delta \rm FUV-NUV}$  &
         $\frac{\Delta \rm [Z/H]}{\Delta \rm FUV-V}$ & $\frac{\Delta \rm [Z/H]}{\Delta \rm FUV-NUV}$ & UV type \\
& (dex mag$^{-1}$) &  (dex mag$^{-1}$) & (\AA\ mag$^{-1}$) & (\AA\ mag$^{-1}$) & (dex mag$^{-1}$) &  (dex mag$^{-1}$) & \\
(1) & (2) & (3) & (4) & (5) & (6) & (7) & (8)\\
\hline
NGC0474 &   0.28    $\pm$   0.15    &   -0.22   $\pm$   0.07    &   2.59    $\pm$   0.72    &   -1.94   $\pm$   0.32    &   0.26    $\pm$   0.30    &   -0.15   $\pm$   0.14  & RSF  \\
NGC0524 &   -0.14   $\pm$   0.05    &   -0.19   $\pm$   0.05    &   -0.40   $\pm$   0.24    &   -0.47   $\pm$   0.25    &   -0.02   $\pm$   0.10    &   0.02    $\pm$   0.10  & UV-weak  \\
NGC0821 &   -0.09   $\pm$   0.07    &   -0.05   $\pm$   0.04    &   -0.49   $\pm$   0.31    &   -0.28   $\pm$   0.17    &   -0.17   $\pm$   0.14    &   -0.09   $\pm$   0.08  & UV-weak    \\
NGC1023 &   -0.79   $\pm$   0.16    &   -0.50   $\pm$   0.07    &   -0.48   $\pm$   0.74    &   -0.18   $\pm$   0.31    &   -0.14   $\pm$   0.31    &   -0.11   $\pm$   0.13  & RSF \\
NGC2695 &   -0.28   $\pm$   0.05    &   -0.26   $\pm$   0.04    &   -0.20   $\pm$   0.22    &   -0.20   $\pm$   0.21    &   -0.04   $\pm$   0.09    &   -0.04   $\pm$   0.08  & UVX  \\
NGC2699 &   -0.24   $\pm$   0.04    &   -0.18   $\pm$   0.03    &   -0.64   $\pm$   0.18    &   -0.49   $\pm$   0.14    &   -0.47   $\pm$   0.08    &   -0.35   $\pm$   0.06  & UV-weak    \\
NGC2768 &   0.04    $\pm$   0.32    &   -0.13   $\pm$   0.14    &   2.39    $\pm$   1.52    &   -1.50   $\pm$   0.66    &   1.21    $\pm$   0.66    &   -0.54   $\pm$   0.29    & UV-weak   \\
NGC2974 &   -0.12   $\pm$   0.04    &   -0.13   $\pm$   0.04    &   -0.46   $\pm$   0.17    &   -0.47   $\pm$   0.18    &   -0.14   $\pm$   0.07    &   -0.14   $\pm$   0.07  & RSF  \\
NGC3032 &   -0.04   $\pm$   0.01    &   -0.09   $\pm$   0.02    &   0.49    $\pm$   0.04    &   1.24    $\pm$   0.10    &   0.26    $\pm$   0.02    &   0.60    $\pm$   0.04   & RSF  \\
NGC4150 &   -0.10   $\pm$   0.02    &   -0.12   $\pm$   0.02    &   0.01    $\pm$   0.08    &   0.01    $\pm$   0.09    &   0.00    $\pm$   0.03    &   0.00    $\pm$   0.04   & RSF \\
NGC4278 &   -0.10   $\pm$   0.04    &   -0.09   $\pm$   0.04    &   -0.87   $\pm$   0.18    &   -0.89   $\pm$   0.17    &   -0.20   $\pm$   0.07    &   -0.21   $\pm$   0.07  & UVX  \\
NGC4374 &   -0.08   $\pm$   0.05    &   -0.08   $\pm$   0.04    &   -0.83   $\pm$   0.24    &   -0.75   $\pm$   0.21    &   -0.22   $\pm$   0.10    &   -0.20   $\pm$   0.09  & UVX  \\
NGC4387 &   -0.12   $\pm$   0.07    &   -0.14   $\pm$   0.04    &   0.40    $\pm$   0.30    &   0.32    $\pm$   0.19    &   0.05    $\pm$   0.12    &   0.00    $\pm$   0.08    & UV-weak  \\
NGC4458 &   -0.26   $\pm$   0.21    &   -0.27   $\pm$   0.11    &   -0.04   $\pm$   0.99    &   0.05    $\pm$   0.53    &   0.58    $\pm$   0.44    &   -0.22   $\pm$   0.27    & UV-weak  \\
NGC4459 &   -0.06   $\pm$   0.01    &   -0.12   $\pm$   0.02    &   -0.17   $\pm$   0.05    &   -0.38   $\pm$   0.10    &   -0.15   $\pm$   0.02    &   -0.30   $\pm$   0.04   & RSF \\
NGC4473 &   0.01    $\pm$   0.15    &   0.03    $\pm$   0.05    &   -1.33   $\pm$   0.68    &   -0.84   $\pm$   0.24    &   -0.54   $\pm$   0.31    &   -0.32   $\pm$   0.11    & UV-weak \\
NGC4477 &   -0.75   $\pm$   0.41    &   -0.55   $\pm$   0.20    &   -0.53   $\pm$   1.92    &   -0.58   $\pm$   0.96    &   -0.44   $\pm$   0.82    &   -0.36   $\pm$   0.41   & UV-weak  \\
NGC4486 &   -0.11   $\pm$   0.04    &   -0.18   $\pm$   0.07    &   -0.60   $\pm$   0.17    &   -1.08   $\pm$   0.31    &   -0.19   $\pm$   0.07    &   -0.33   $\pm$   0.12   & UVX \\
NGC4526 &   -0.08   $\pm$   0.03    &   -0.17   $\pm$   0.10    &   0.08    $\pm$   0.16    &   -0.53   $\pm$   0.45    &   -0.15   $\pm$   0.07    &   0.35    $\pm$   0.18   & RSF  \\
NGC4546 &   -1.03   $\pm$   0.23    &   -0.33   $\pm$   0.07    &   -3.03   $\pm$   1.10    &   -0.93   $\pm$   0.33    &   -1.03   $\pm$   0.52    &   -0.33   $\pm$   0.16   & UV-weak   \\
NGC4550 &   0.44    $\pm$   0.11    &   0.26    $\pm$   0.06    &   -0.66   $\pm$   0.48    &   -0.43   $\pm$   0.27    &   -0.64   $\pm$   0.22    &   -0.37   $\pm$   0.13   & RSF \\
NGC4552 &   -0.06   $\pm$   0.03    &   -0.09   $\pm$   0.04    &   -0.34   $\pm$   0.12    &   -0.49   $\pm$   0.17    &   -0.09   $\pm$   0.05    &   -0.13   $\pm$   0.07   & UVX \\
NGC4564 &   -0.07   $\pm$   0.03    &   -0.08   $\pm$   0.04    &   -0.30   $\pm$   0.14    &   -0.35   $\pm$   0.17    &   -0.09   $\pm$   0.06    &   -0.10   $\pm$   0.08   & UVX \\
NGC4570 &   -0.31   $\pm$   0.07    &   -0.23   $\pm$   0.05    &   -0.60   $\pm$   0.31    &   -0.40   $\pm$   0.24    &   -0.14   $\pm$   0.15    &   -0.11   $\pm$   0.12   & UV-weak   \\
NGC4621 &   -0.08   $\pm$   0.03    &   -0.09   $\pm$   0.03    &   -0.48   $\pm$   0.12    &   -0.49   $\pm$   0.13    &   -0.16   $\pm$   0.06    &   -0.17   $\pm$   0.06   & UVX  \\
NGC5198 &   -0.15   $\pm$   0.04    &   -0.15   $\pm$   0.04    &   -0.39   $\pm$   0.19    &   -0.39   $\pm$   0.21    &   -0.13   $\pm$   0.09    &   -0.14   $\pm$   0.09   & UVX \\
NGC5308 &   -0.41   $\pm$   0.09    &   -0.33   $\pm$   0.07    &   -0.76   $\pm$   0.44    &   -0.60   $\pm$   0.34    &   -0.31   $\pm$   0.20    &   -0.24   $\pm$   0.15   & UV-weak   \\
NGC5813 &   0.04    $\pm$   0.15    &   0.07    $\pm$   0.08    &   0.10    $\pm$   0.70    &   -0.10   $\pm$   0.36    &   0.59     $\pm$   0.30    &   0.23    $\pm$   0.15    & A/Z \\
NGC5831 &   -0.12   $\pm$   0.06    &   -0.08   $\pm$   0.04    &   -1.33   $\pm$   0.30    &   -1.20   $\pm$   0.20    &   -0.20   $\pm$   0.13    &   -0.19   $\pm$   0.09   & A/Z \\
NGC5838 &   -0.27   $\pm$   0.04    &   -0.28   $\pm$   0.04    &   -0.50   $\pm$   0.18    &   -0.52   $\pm$   0.18    &   -0.14   $\pm$   0.07    &   -0.15   $\pm$   0.08   & UV-weak   \\
NGC5845 &   -0.17   $\pm$   0.05    &   -0.11   $\pm$   0.03    &   -1.44   $\pm$   0.23    &   -0.96   $\pm$   0.16    &   -0.30   $\pm$   0.10    &   -0.20   $\pm$   0.07   & A/Z   \\
NGC5846 &   -0.03   $\pm$   0.05    &   -0.04   $\pm$   0.06    &   -0.54   $\pm$   0.22    &   -0.65   $\pm$   0.28    &   -0.19   $\pm$   0.09    &   -0.22   $\pm$   0.11   & UVX  \\
NGC5982 &   -0.14   $\pm$   0.04    &   -0.13   $\pm$   0.04    &   -0.59   $\pm$   0.20    &   -0.59   $\pm$   0.20    &   -0.20   $\pm$   0.08    &   -0.21   $\pm$   0.08  & UVX   \\
NGC7457 &   -0.02   $\pm$   0.03    &   -0.02   $\pm$   0.03    &   0.02    $\pm$   0.14    &   0.02    $\pm$   0.12    &   0.02    $\pm$   0.05    &   0.02     $\pm$   0.04   & RSF \\
\hline
\end{tabular}

Columns: (1) Galaxy identifier. (2)--(7): Gradients of the UV
colour-index relations. (8): UV--optical radial colour profiles
classification. RSF: recent star formation. UVX: UV upturn.
\end{table*}

%
%
\begin{figure*}
\begin{center}
\includegraphics[width=0.85\textwidth,clip=m]{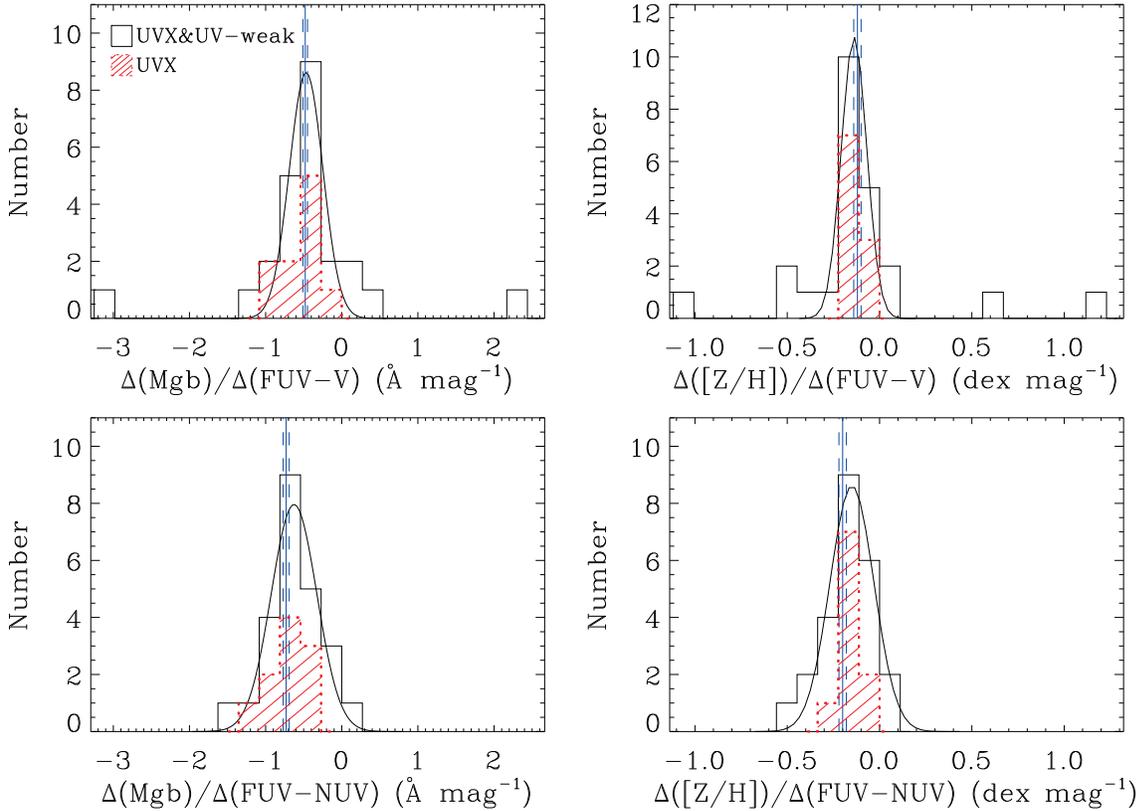}
\end{center}
\caption[]{Histograms of the individual gradients of the UV
colour--line index relations for the sample galaxies. The black
solid lines are Gaussian fits to all quiescent galaxies. The blue
vertical solid lines show the global gradients determined from
fitting the luminosity-weighted galaxy profiles (solid lines in
Figures~\ref{fig:fuvv} and \ref{fig:fnuv}), while the dashed lines
indicate the associated uncertainties.} \label{fig:histo1}
\end{figure*}

%
%
\begin{figure*}
\begin{center}
\includegraphics[width=0.85\textwidth,clip=m]{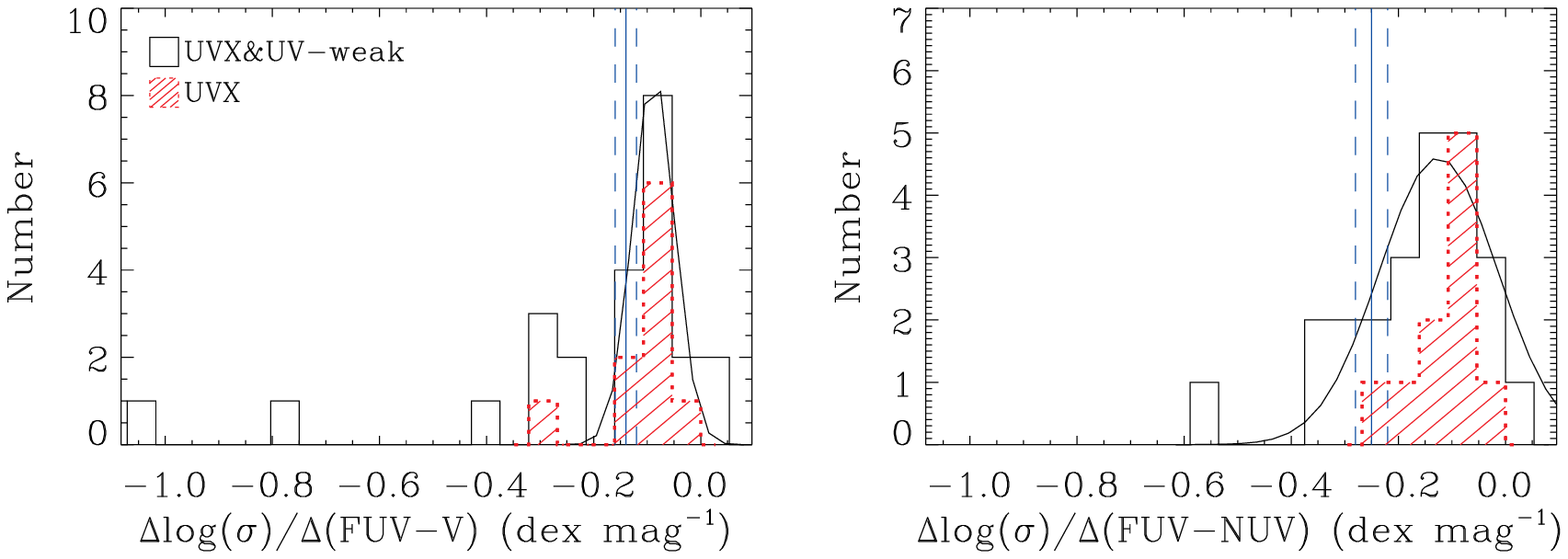}
\end{center}
\caption[]{Same as Figure~\ref{fig:histo1} but for the UV
colour--stellar velocity dispersion relations. The blue vertical
solid lines show the global gradients determined from fitting the
luminosity-weighted galaxy profiles (solid lines in
Figure~\ref{fig:vd}).} \label{fig:vd_histo2}
\end{figure*}

Most importantly, only \mgb\ and [Z/H] internal gradients for
quiescent galaxies show correlations. This suggests that either
Mg is a primary driver of the UV upturn, given that \mgb\ is a
reliable tracer of Mg, or Mg and the UV upturn share the same
primary driver. If the  stellar population properties
([$\alpha$/Fe], [Z/H] and age) are reliable, we conclude further
that the UV strength is not predominantly governed by the
$\alpha$-enhancement. The correlation between the [Z/H] index and
the UV colour can be attributed to the role of Mg in the [Z/H]
measurement. The weak or absent correlation with
$\alpha$-enhancement ([$\alpha$/Fe]), however, is contrary to our
expectations.

Analogously to Figure~\ref{fig:fuvv_grad},
Figure~\ref{fig:fnuv_grad} shows the FUV$-$NUV logarithmic internal
radial colour gradients versus the internal gradients of the line
indices and stellar population properties. The (FUV$-$NUV)--\mgb\
correlation is again much tighter than the (FUV$-V$)--\mgb\
correlation, as discussed in Section~\ref{sec:radial}, and the
correlation with \mgb\ is much stronger than that with [Z/H].

Figure~\ref{fig:vd_grad} shows the internal gradients of the
velocity dispersion as a function of the UV internal colour
gradients. Comparing to Figures~\ref{fig:fuvv_grad} and
\ref{fig:fnuv_grad}, the correlations with $\sigma_{\rm e}$ are much
weaker than those with both \mgb\ and [Z/H]. This suggests that the
UV upturn is driven by stellar population properties rather than the
velocity dispersion.

For clarity, we divide galaxies into groups exhibiting similar UV
radial colour profiles for the region interior to the effective
radius (see Figure~1 of \citeauthor{jetal09}) and listed in
Table~\ref{tab:grad}. This is unlike the classification of
\citeauthor{jetal09}, that considered much larger radial ranges,
because \sauron\ covers each galaxy only up to about one effective
radius. The classification differences between \citeauthor{jetal09}
and this paper thus arise from different radial ranges considered.
To select purely old galaxies, we first exclude galaxies with recent
star formation as listed in Table~1 of \citeauthor{jetal09}.
Galaxies included in this category are NGC~474, 1023, 2974, 3032,
4150, 4459, 4526, 4550 and 7457. We then consider the NUV$-V$ radial
colour profiles. As discussed in Section~3.3 of
\citeauthor{jetal09}, NUV$-V$ colours usually show negative slopes
due to the gradients of the underlying stellar populations, mainly
in age and metallicity. This age-metallicity degeneracy is
well-known and it is difficult to distinguish the effects of a small
change in age from those of a small change in metallicity.
Therefore, we classify a galaxy as an age-metallicity degeneracy
galaxy if its NUV$-V$ profile is below 5.0~mag~arcsec$^{-2}$ around
the effective radius only. These galaxies include NGC~5813, 5831 and
5845. Secondly, we consider the FUV$-$NUV and FUV$-V$ profiles
inside the effective radius. If a galaxy has a central region with
FUV$-$NUV$\,<\,0.9$~mag~arcsec$^{-2}$ and
FUV$-V\,<\,6.21$~mag~arcsec$^{-2}$ then, following the new
classification scheme of \citet{yetal11}, we classify it as a UV
upturn galaxy. These galaxies are NGC~2695, 4278, 4374, 4486, 4552,
4564, 4621, 5198, 5846 and 5982. Finally, we classify the remaining
galaxies as UV-weak galaxies. Galaxies included in this category are
NGC~524, 821, 2699, 2768, 4387, 4458, 4473, 4477, 4546, 4570, 5308
and 5838.

Furthermore, we calculate the gradients of the UV colour--index
relations (e.g. $\Delta$(\mgb)/$\Delta$(FUV$-V$)) for each
individual galaxy using least-square fits (see Table~\ref{tab:grad})
in the same way as described in Section~\ref{sec:gradient}. We note
that the values of e.g. [$\Delta$(\mgb)/$\Delta$(FUV$-V$)] are not
equal to those of
[$\Delta$(\mgb)/$\Delta$(log\,(R))]/[$\Delta$(FUV$-V$)/$\Delta$(log\,(R))]
listed in Table~\ref{tab:rgrad}.

In Figure~\ref{fig:histo1}, we show the histograms of the \mgb\ and
[Z/H] gradients with respect to the UV colours of individual
galaxies (e.g. $\Delta$(\mgb)/$\Delta$(FUV$-V$)). We present  the
cases of \mgb\ and [Z/H] only, showing the tightest correlations in
Figures~\ref{fig:fuvv_grad} and ~\ref{fig:fnuv_grad}. These diagrams
are useful to test whether the individual galaxy gradients (local
gradients) follow the global correlations (solid lines in
Figures~\ref{fig:fuvv} and \ref{fig:fnuv}). The black histograms
represent UV-weak and UV upturn galaxies, whereas the red shaded
histograms show UV upturn galaxies only. We also fit Gaussians to
the UV-weak and UV upturn galaxy gradient distributions and plot
them as black solid lines in each panel. The blue vertical solid
lines show the global gradients (solid lines in
Figures~\ref{fig:fuvv} and \ref{fig:fnuv}), while the blue dashed
lines indicate the associated uncertainties.

If the global correlations (i.e. between galaxies) also apply
locally within galaxies, the means of the Gaussian fits should be
consistent with the global gradients. As discussed in
Section~\ref{sec:radial}, our global correlations are consistent
with those of \citeauthor{betal11} (determined from a fit to $R_{\rm
e}$/2 aperture values for quiescent galaxies). For example, the
slopes of the best-fit (FUV$-$NUV)$-$\mgb, (FUV$-$NUV)$-$Fe5015 and
(FUV$-$NUV)$-$\hb\ relations based on integrated properties are
-0.64$\pm$0.12, -0.21$\pm$0.17 and 0.20$\pm$0.08 \AA\ mag$^{-1}$,
whereas our global slopes are -0.73$\pm$0.04, -0.36$\pm$0.06 and
-0.19$\pm$0.02  \AA\ mag$^{-1}$, respectively. We note that the
distributions of the individual galaxy gradients are consistent with
the global gradients within the errors, although the distributions
are considerably broader (Figure~\ref{fig:histo1}). For example, the
global gradient for (FUV$-V$)--\mgb\ is
$-0.48\pm0.03$~\AA~mag$^{-1}$, whereas the mean of the Gaussian fit
to the local gradient distribution is $-0.46$~\AA~mag$^{-1}$, with
an rms of $0.21$~\AA~mag$^{-1}$. Comparing the black and red shaded
histograms, one can see that the distributions of the UV upturn
galaxies are narrower than those of the UV-weak galaxies.

Similarly to Figure~\ref{fig:histo1}, Figure~\ref{fig:vd_histo2}
shows the histograms of the stellar velocity dispersion gradients
with respect to the UV colours of UV-weak and UV upturn galaxies. In
the case of the velocity dispersion, however, the global gradients
(vertical solid lines) are different from the local gradients (i.e.
individual galaxies). This implies that the UV strength is not
highly dependent on structural or dynamical properties.

%
%
\begin{figure*}
\begin{center}
\includegraphics[width=13cm]{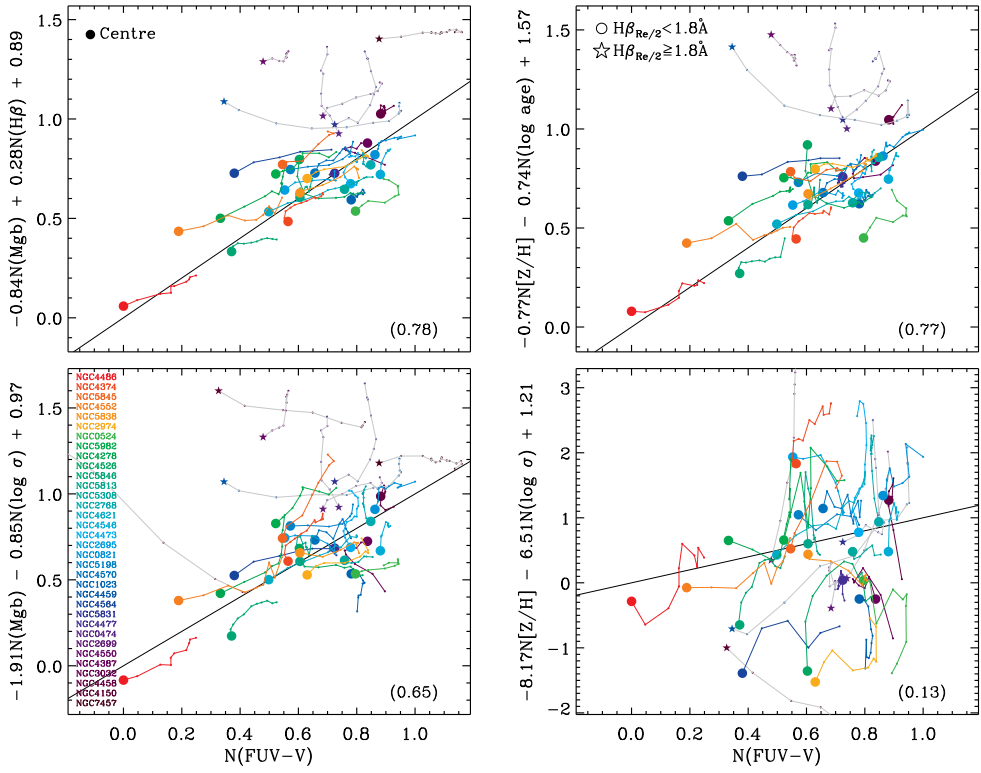}
\end{center}
\caption[]{Edge-on views of the best-fit planes for selected
three-parameter relations. The correlation coefficient is given in
the bottom-right corner of each panel. The multi-parameter
correlations shown are not much tighter than the simple
one-parameter fits shown in Figures~\ref{fig:fuvv} and
\ref{fig:fnuv}.} \label{fig:fuvv_3d}
\end{figure*}
%

%
%
\begin{figure*}
\begin{center}
\includegraphics[width=13cm]{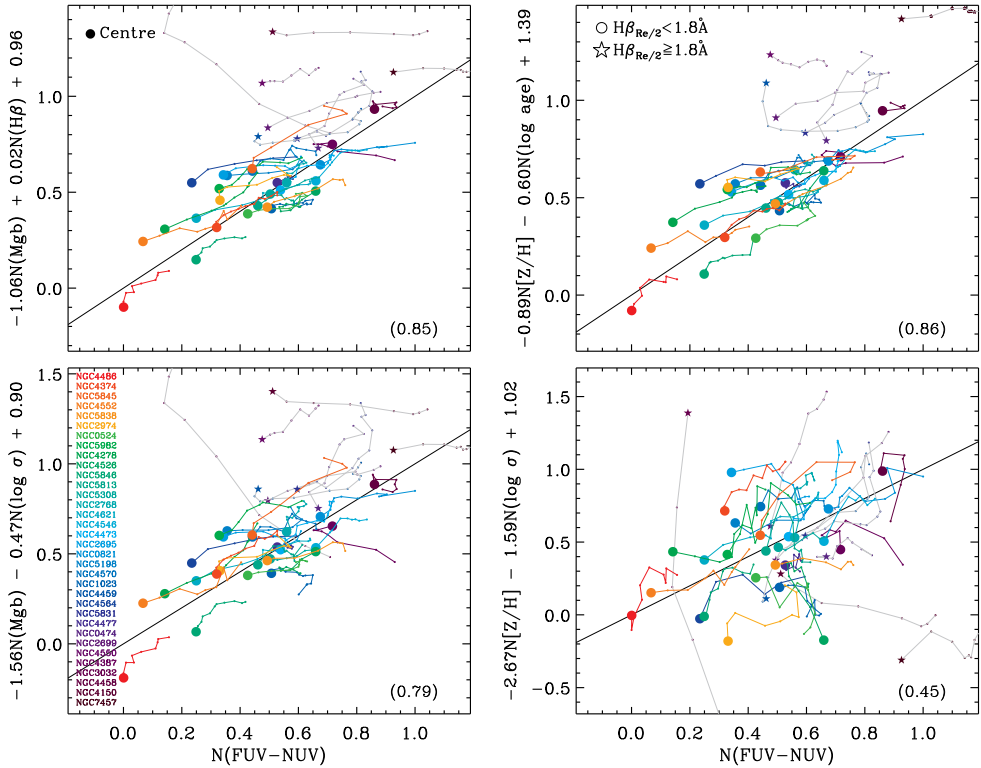}
\end{center}
\caption[]{Same as Figure~\ref{fig:fuvv_3d} but for the FUV$-$NUV
colour.} \label{fig:fnuv_3d}
\end{figure*}
\subsection{Multi-dimensional relations}
\label{sec:multi}
Figures~\ref{fig:fuvv}--\ref{fig:fnuv}, \ref{fig:fuvv_grad} and
\ref{fig:fnuv_grad} show that the (FUV$-$NUV)--\mgb\ relation is the
tightest correlation. We now consider multiple-parameter relations
analogous to the fundamental plane \citep{dd87,detal87}.
Three-parameter relations are not trivially expected in the absence
of {\em a priori} physical foundations, such as the virial theorem
in the case of the Fundamental Plane. To minimise biases arising
from the very different definitions and hence dynamic ranges of the
parameters considered (e.g.  line strengths,
 stellar population properties, velocity dispersion and
colours), we normalize each quantity to yield an observed range
of $1.0$. For example, the observed range of FUV$-$V is roughly
$4.7$ to $7.3$ mag (a difference of $2.6$ magnitudes), and so in
the abscissa of Figure~\ref{fig:fuvv_3d}, $0.0$ means $2.6$ mag
bluer (UV stronger) than $1.0$. Likewise, for the ordinate of the
bottom-left plot, the observed range of \mgb\ is roughly $3.2$ to
$5.3$\,\AA\ (a difference of $2.1$\,\AA), and that of
log~($\sigma$/km\,s$^{-1}$) is roughly $1.9$ to $2.5$
(a difference of $0.6$).

We have explored a number of parameter combinations and present some
cases showing tight correlations in Figures~\ref{fig:fuvv_3d} and
\ref{fig:fnuv_3d}.

The correlation coefficients reported in the bottom-right corner of
each panel in Figures~\ref{fig:fuvv_3d} and \ref{fig:fnuv_3d} are
not much better (i.e. larger) than the simple one-parameter fits
shown in Figures~\ref{fig:fuvv} and \ref{fig:fnuv}. This means that
the relative FUV strength is not significantly better described when
a third parameter is employed. This is in agreement with our finding
in \citeauthor{betal11}, that it is stellar population properties
rather than galaxy dynamical properties that govern the UV strength.

%
%
\section{THE UV--INDEX RELATIONS}
\label{sec:discussion1}

In summary, the results presented in Section~\ref{sec:results}
indicate tight local and global correlations between UV colours and
\mgb\, and marginal correlations between UV colours and [Z/H]. These
UV--\mgb\ and UV--[Z/H] relations are all the more significant
because the global gradients are consistent with the local
gradients. If the UV strength is related to the Mg line strength,
galaxies with larger Mg gradients should have larger UV colour
gradients, and the global gradient across galaxies should be
consistent with the local gradients of individual galaxies as
observed. Therefore, our results suggest that Mg is a main driver of
the UV upturn or Mg and the UV upturn share the same primary driver,
and the correlations between UV colours and [Z/H] can be understood
by the role of Mg in the [Z/H] measurement.

%
%
\begin{figure*}
\begin{center}
\includegraphics[width=0.85\textwidth,clip=]{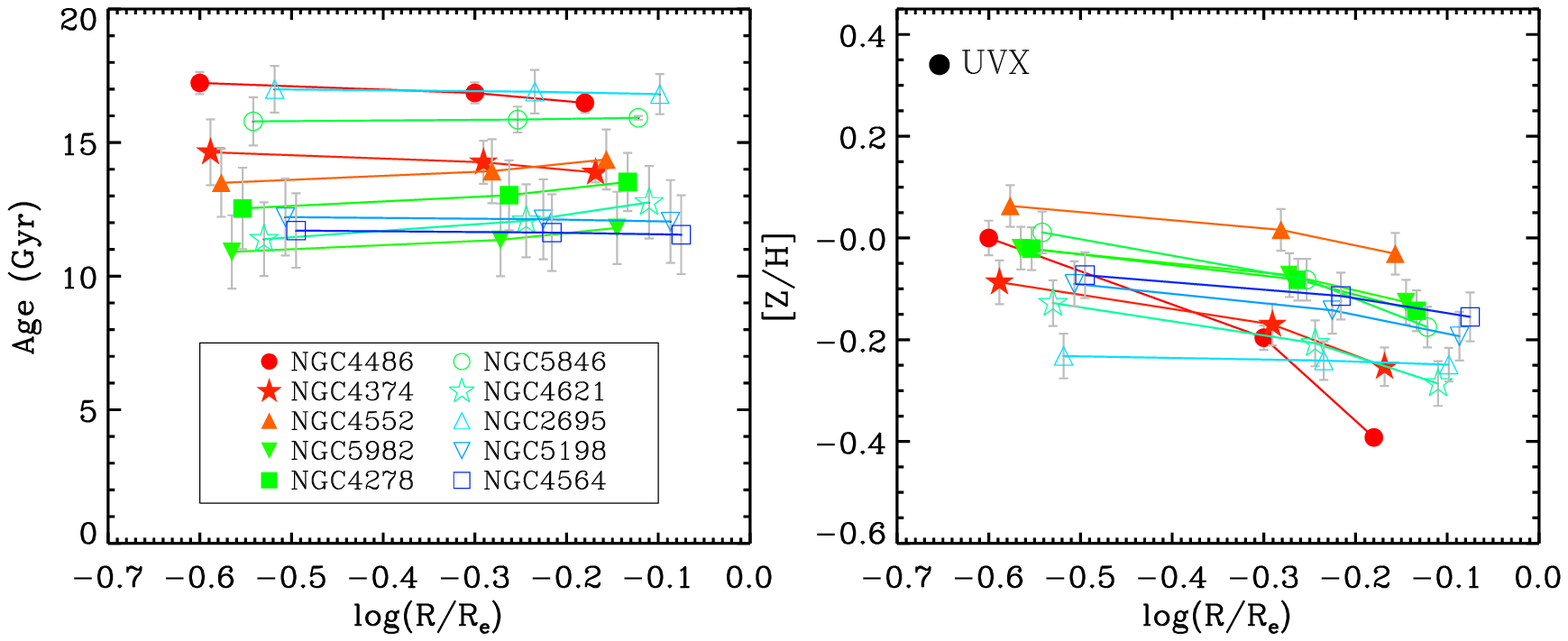}
\end{center}
\caption[]{Radial profiles of age and metallicity ([Z/H]) for UV
upturn galaxies. The values are derived from NUV-defined elliptical
annuli at $R_{\rm e}/4$, $R_{\rm e}/2$ and 2$R_{\rm e}/3$ . Each
profile is given a slightly different $x$-axis offset for clarity. }
\label{fig:sp_uvx}
\end{figure*}
%

%
%
\begin{figure*}
\begin{center}
\includegraphics[width=0.85\textwidth,clip=]{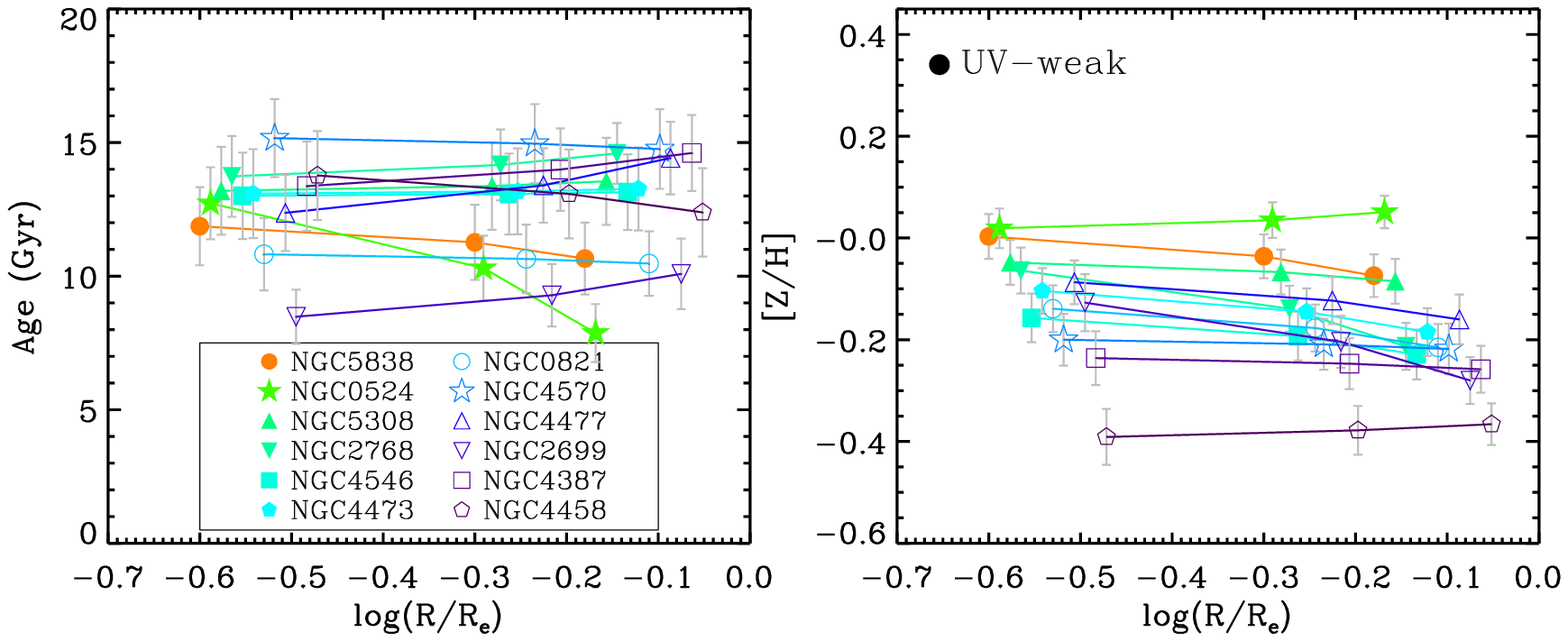}
\end{center}
\caption[]{Same as Figure~\ref{fig:sp_uvx} but for UV-weak
galaxies.} \label{fig:sp_red}
\end{figure*}

Mg being the most representative $\alpha$ element, it is puzzling
that [$\alpha$/Fe] does not show a clear correlation with UV.
Meanwhile, it should be noted that \citet{cetal11} found good
correlations between integrated FUV$-$NUV colours and integrated
[$\alpha$/Fe] and [Z/H]. The weak correlations between the FUV
colours and  [$\alpha$/Fe] in  this paper may be due to the fact
that the $\alpha$-element is not directly measured from line
strengths but derived from model fitting and so much more difficult
than measuring a single metal line (e.g. \mgb). Furthermore,
the stellar population properties ([$\alpha$/Fe], [Z/H] and age)
depend on the stellar models and spectral library adopted \citep[see
e.g.][]{coetal07, petal09}.

Given the reasonably-clear correlation between [Z/H] and UV colours,
the apparently-weak correlation shown by Fe5015 and UV colours seems
mysterious. A possibility is that Fe5015 may not be a robust tracer
of [Z/H] because it is heavily contaminated by poorly understood
absorption lines in this band \citep{leeetal09}.

We have also shown that global correlations exist between UV-related
colours and the \hb\ line strength, but we find no correlation
between UV-related colour gradients and internal \hb\ gradients.
This can also be explained by metallicity effects on \hb\ as
discussed in Section 3.1.

Given the above, the metallicity gradients of UV upturn galaxies
must be preserved to the present day. Is this possible in the
hierarchical paradigm? \citet{hetal09} carried out simulations and
claimed that metallicity gradients are only weakly affected by both
wet and dry mergers. \citet{detal09} also argued that metallicity
gradients can be preserved in major dry mergers if the metallicity
gradient of companion is sufficiently steep.

Lastly, we point out that galaxies having recent star formation are
outliers in the UV--\mgb\ and UV--[Z/H] relations; they usually have
lower \mgb\ and [Z/H] values. It is therefore clear that tight
UV--\mgb\ and UV--[Z/H] relations are only present for quiescent
galaxies.

\section{ORIGIN OF THE UV UPTURN}
\label{sec:discussion2}
\subsection{Stellar population parameters from line strengths}
\label{sec:mod_line}

If different UV strengths are caused by different stellar population
properties, this implies a difference between the stellar
populations of UV upturn and UV-weak galaxies. To attempt to
understand the conditions that create the observed UV upturns, we
explore the  stellar population parameters after separating
quiescent galaxies into UV upturn and UV-weak galaxies, based on
their UV radial colour profiles as discussed in Section~3.3.

%
%
\begin{figure*}
\begin{center}
\includegraphics[width=0.85\textwidth,clip=]{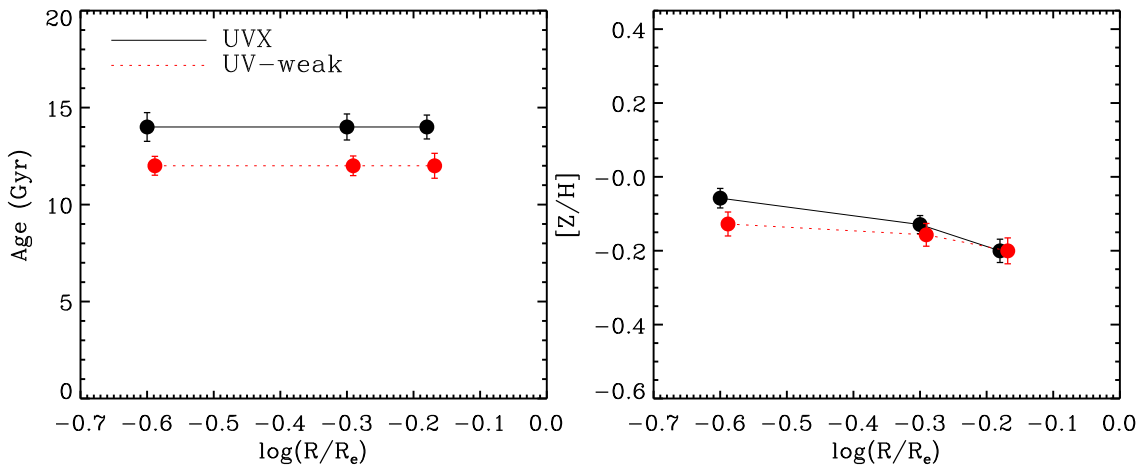}
\end{center}
\caption[]{The mean radial age and metallicity profiles of UV upturn
(black solid lines) and UV-weak (red dotted lines) galaxies based on
the derived stellar population parameters from
\citeauthor{ketal10}.} \label{fig:sp_all}
\end{figure*}

Figure~\ref{fig:sp_uvx} shows the age and [Z/H] values of UV upturn
galaxies at three different radii. To derive these stellar
population parameters, we used here NUV-defined elliptical annuli of
semi-major axes  $R_{\rm e}/4$  (i.e. central), $R_{\rm e}/2$ (i.e.
intermediate) and 2$R_{\rm e}/3$  (i.e. outskirts), respectively.
For clarity, each profile is given a slightly different $x$-axis
offset. In UV upturn galaxies, the {\it derived} ages extend from
roughly $11$ to $17$~Gyr. In the case of [Z/H], the range is roughly
$-0.25$ to $+0.05$. We note that the age gradients are generally
flat, whereas significant metallicity gradients exist.

Similarly to Figure~\ref{fig:sp_uvx}, Figure~\ref{fig:sp_red} shows
the age and [Z/H] gradients of UV-weak galaxies. The age gradients
are again nearly flat, even though the metallicity gradients are
weaker than those of UV upturn galaxies.

For comparison, we construct the mean radial age and metallicity
profiles for the samples of UV upturn and UV-weak galaxies. The
results are shown in Figure~\ref{fig:sp_all}; the errors correspond
to the standard deviations about the mean. There is a notable
difference in age between the two samples (left panel), and a slight
[Z/H] difference (right panel), especially in the central regions.
Based on this, one might think that it is the age difference that
governs the UV strength. However, without a radial age gradient, age
alone cannot explain the central concentration of FUV light
observed. Metallicity gradients are steep, especially in UV upturn
galaxies, and so are probably essential to explain the
centrally-concentrated FUV excess (as suggested in
Sections~\ref{sec:results} and \ref{sec:discussion1}).

\subsection{New stellar population parameters from colours}
\label{sec:mod_colour}

If it is indeed stellar population properties that govern the UV
strength of a quiescent galaxy, as we concluded in
Sections~\ref{sec:results} and \ref{sec:discussion1}, then the same
stellar population models should be able to reproduce the UV
relations for both integrated and internal properties. To find the
range of age and metallicity that reproduce the observed UV
properties, we attempt to fit the observed colour profiles (rather
than line strengths) with the stellar population synthesis
models of \citet{yi03} and \citet{ydo97, ydo98} that were calibrated
to reproduce the observed properties of the UV upturn phenomenon.
These models are sensitive to metallicity, age, stellar mass loss,
helium enrichment ($\Delta$Y/$\Delta$Z) and the distribution of
stars on the HB \citep*[see e.g.][]{dor95, ps11,cyl11}.

\begin{figure*}
\begin{center}
\includegraphics[width=0.85\textwidth,clip=]{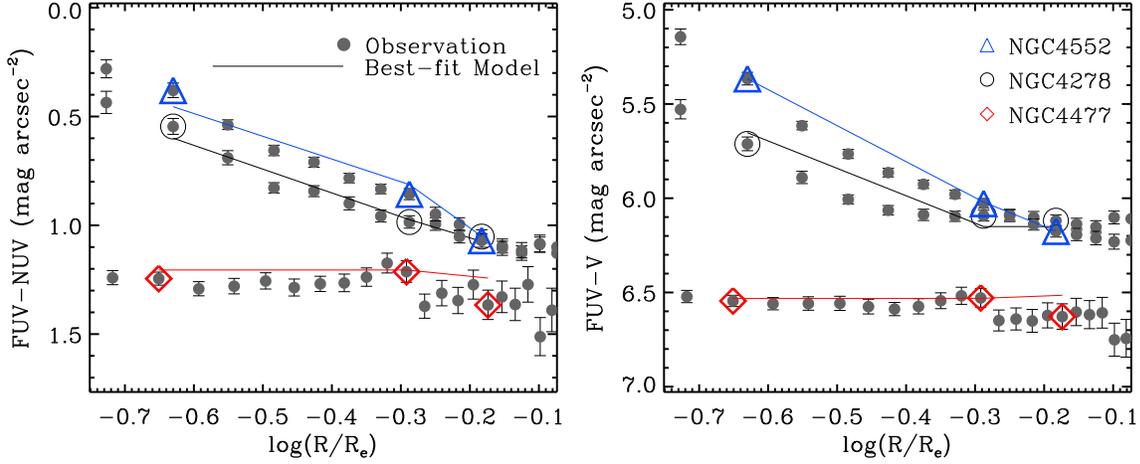}
\end{center}
\caption[Comparison between the observed magnitudes and metal-rich
models.]{Comparison between the observed colours and metal-rich
models of three sample galaxies (NGC\,4552, 4278 and 4477). The
FUV$-$NUV ({\it left}) and FUV$-$V ({\it right}) colours of each
galaxy are shown as a function of radius. To derive the age and
metallicity at a given radius, we fit the observed colours with the
stellar population synthesis models of \citet{yi03}. Three different
regions for each galaxy are used (open symbols; center, intermediate
radius and outskirt). The best-fit model is shown as a solid line.}
\label{fig:mfit1}
\end{figure*}

\begin{figure*}
\begin{center}
\includegraphics[width=0.85\textwidth,clip=]{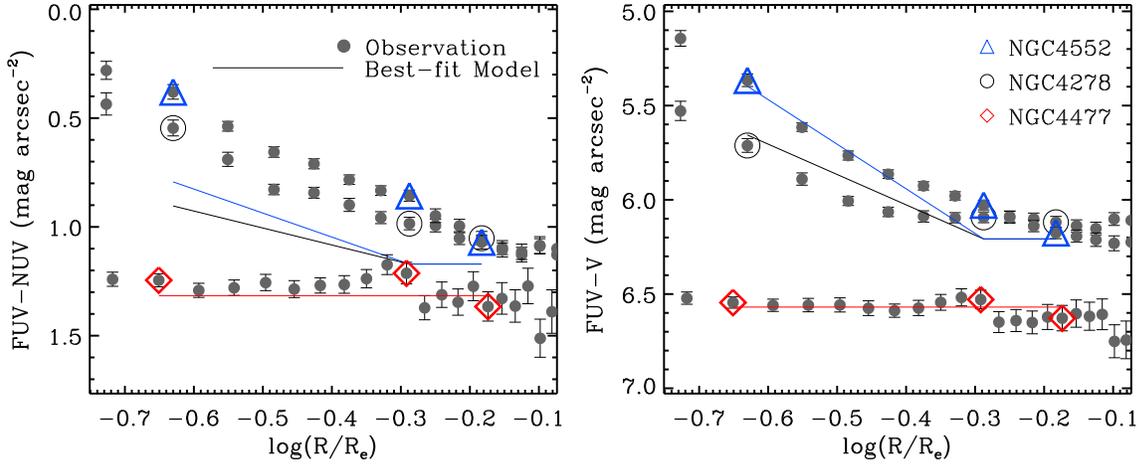}
\end{center}
\caption[Same as Figure~\ref{fig:mfit1} but for the metal-poor
model.]{Same as Figure~\ref{fig:mfit1} but for the metal-poor
model.} \label{fig:mfit2}
\end{figure*}

The comparisons between the observed colours and models are shown in
Figures~\ref{fig:mfit1} and \ref{fig:mfit2}. The existence of strong
radial gradients of UV colours within galaxies is well known
\citep[e.g.][]{o92, oetal98}, and in early-type galaxies they are
likely the result of underlying stellar population gradients, mainly
in age or metallicity or both. Therefore, we consider two simple
cases: (1) an age gradient and (2) a metallicity gradient.

%
%
%

%
%

In the case of the metal-rich model, suggesting that the metal-rich
HB stars are the most likely UV sources (see
Section~\ref{sec:intro}; \citealt{gr90,hdp92,bcf94,dor95,ydk97}), we
could not reproduce the observed colour gradients at all using an
age gradient alone. However, if we consider an internal metallicity
gradient (with no age gradient), metal-rich models are able to
reproduce the observed colour profiles. We assume, therefore, that
the internal colour gradients of quiescent early-type galaxies are
the result of simple metallicity gradients. The overall age of an
individual galaxy is permitted to vary over the range 1 to 20~Gyr.

To constrain the age and metallicity of each galaxy, we thus fit the
model to the observed FUV, NUV and $V$ surface brightnesses at the
same three radii used previously (center ($R_{\rm e}/4$),
intermediate radius ($R_{\rm e}/2$) and outskirts ($2R_{\rm e}/3$)),
and compute the associated $\chi^2$ statistics to obtain a
probability distribution of the age and metallicity.
Figure~\ref{fig:mfit1} shows the observed FUV$-$NUV and FUV$-V$
colour profiles (filled symbols) and the best-fit model profiles
(solid lines) of NGC\,4552 (blue triangle), 4278 (black circle) and
4477 (red diamond). Three open symbols of each galaxy represent the
three fitted regions. This figure shows that we have found a
plausible range of metal-rich model parameters that is capable of
reproducing the observed amplitude of the UV upturn with the
conventional values of $\Delta Y/\Delta Z$~=~2.5, $\eta$~=~0.7 and
[$\alpha$/Fe]~=~0.0.

Metal-poor models, suggesting that the dominant UV sources are older
and metal-poor HB stars and their post-HB progeny (see
Section~\ref{sec:intro};  \citealt{ldz94,pl97,bg08}), fail to
reproduce the observed surface brightnesses of UV upturn galaxies
(e.g. NGC~4552 and 4278) when we consider age and metallicity
gradients, as shown in Figure~\ref{fig:mfit2}, although this depends
on the adopted models \citep[see e.g.][]{pl97}. We note that the
model-fit for UV weak galaxy (e.g. NGC~4477) based on metal-poor
models is also reasonably good. A number of recent studies found
peculiar globular clusters having EHB stars that can be explained by
the presence of super-helium-rich populations \citep{letal05,
lgc07}, which could also be a copious  source of FUV flux. 
Indeed, \citet{cyl11} claimed that their models with helium-enhanced
subpopulations can reproduce the observed spectra of NGC\,4552 and
4649. To perform the same test for this metal-poor but helium
enhanced hypothesis, however, require super-helium-rich stellar
population models which are not fully available yet.


%
%
\begin{figure*}
\begin{center}
\includegraphics[width=0.85\textwidth,clip=]{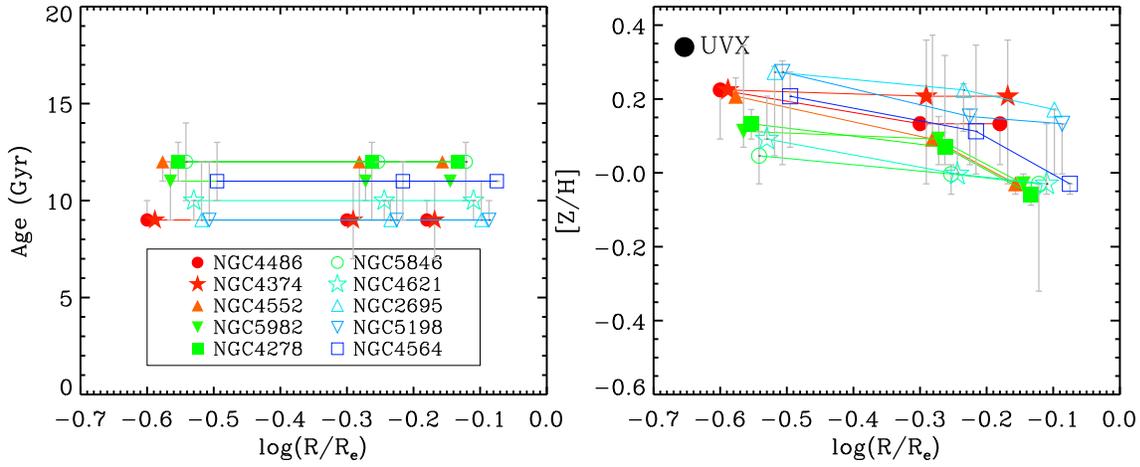}
\end{center}
\caption[]{Radial profiles of age and metallicity for UV upturn
galaxies, as derived from fits to the observed colours and the
stellar population synthesis models of \citet{yi03} at $R_{\rm
e}/4$, $R_{\rm e}/2$ and 2$R_{\rm e}/3$. Each profile is given a
slightly different $x$-axis offset for clarity.} \label{fig:mod_uvx}
\end{figure*}
%

%
%
\begin{figure*}
\begin{center}
\includegraphics[width=0.85\textwidth,clip=]{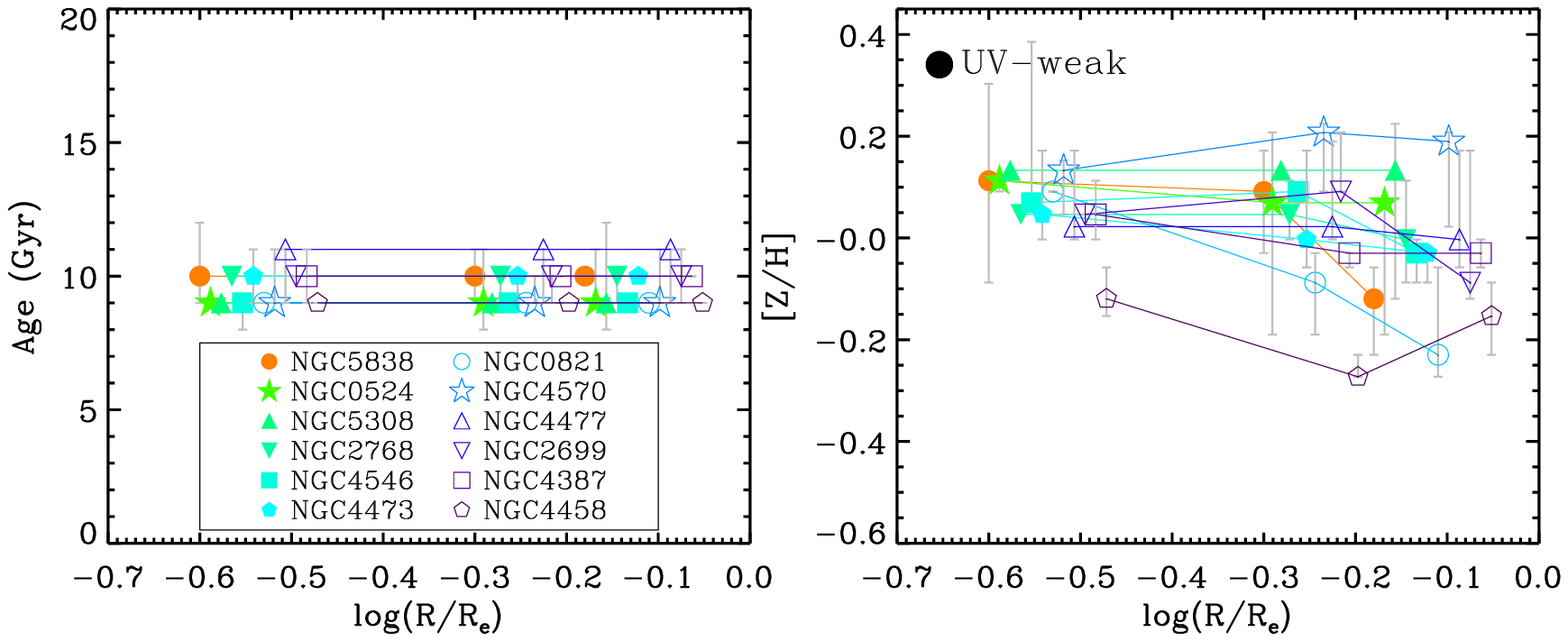}
\end{center}
\caption[]{Same as Figure~\ref{fig:mod_uvx} but for UV-weak
galaxies.} \label{fig:mod_red}
\end{figure*}
%

%
%
\begin{figure*}
\begin{center}
\includegraphics[width=0.85\textwidth,clip=]{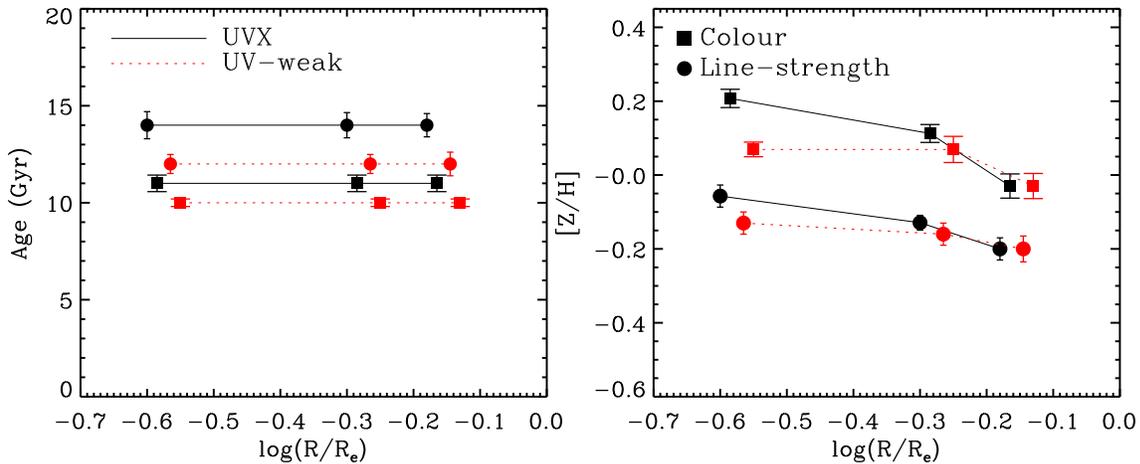}
\end{center}
\caption[]{The mean radial age and metallicity profiles of UV upturn
(black solid lines) and UV-weak (red dotted lines) galaxies based on
colour fits and the models of \citet{yi03} (filled squares). For
comparison, we also show the mean ages and metallicities of UV
upturn and UV-weak galaxies based on the derived stellar population
parameters of \citeauthor{ketal10} (filled circles; see
Figure~\ref{fig:sp_all}).} \label{fig:mod_all}
\end{figure*}

Analogously to Figure~\ref{fig:sp_uvx}, the best-fit values of age
and metallicity based on metal-rich models, for all galaxies
classified as UV upturn from their radial colour profiles, are shown
in Figure~\ref{fig:mod_uvx}. For clarity, the profiles are again
given slight offsets with respect to each other along the $x$-axis.
Errors show parameter values within 1$\sigma$ of the best-fit model.
According to these models, the age range is 9 to 12~Gyr and the
metallicity range is roughly $-0.05$ to $+0.28$. Similarly to
Figures~\ref{fig:sp_red} and \ref{fig:mod_uvx},
Figure~\ref{fig:mod_red} shows best-fit values of age and
metallicity for UV-weak galaxies. In this case, the range of ages is
9 to 11~Gyr, with metallicities of roughly $-0.27$ to $+0.20$.

Like Figure~\ref{fig:sp_all}, Figure~\ref{fig:mod_all} shows the
mean stellar population properties and the standard deviations about
the means for UV upturn and UV-weak galaxies (filled squares).
According to the models, the mean stellar age of UV-upturn galaxies
(black solid lines) is about 11~Gyr and their central regions have a
metallicity distribution peaking around $0.2$, while at the
effective radius the metallicities decrease to slightly sub-solar.
UV-weak galaxies (red dotted lines) have a similar mean stellar age
but significantly smaller metallicity gradients compared to UV
upturn galaxies. The metallicity difference between UV upturn and
UV-weak galaxies in the central regions produces the difference in
the UV upturn strength. The UV upturn phenomenon seems indeed to be
due to stellar population properties.

For comparison, we plot again the mean ages and metallicities
derived from spectral line strengths (see Figure~\ref{fig:sp_all})
as filled circles. In both cases, UV upturn galaxies are slightly
older than UV-weak galaxies and the metallicities of UV upturn and
UV-weak galaxies reveal differences, especially in the central
regions. The trends of age and metallicity derived from spectral
line strengths and colours are thus similar, but the actual values
reveal significant differences. This can be explained in terms of
the age-metallicity degeneracy.

The previous \sauron\ work of \citeauthor{ketal10} found that models
with very high ages ($\approx$14~Gyr) match the high Mg absorption
of UV upturn galaxies. In contrast, through our colour fitting
method, we find that slightly younger ($\approx$11~Gyr) but more
metal-rich models better reproduce the pronounced UV flux in the
interiors of UV upturn galaxies. It is well known that line
strengths and colour indices suffer from the age-metallicity
degeneracy (Worthey 1994). Furthermore, as described in
Section~\ref{sec:sauron}, we should not believe the absolute ages
and metallicities returned by the models, even though they still
allow us to make relative comparisons between galaxies. The model
parameters derived from line strengths in \citeauthor{ketal10}
(Section~\ref{sec:mod_line}) and from colours here
(Section~\ref{sec:mod_colour}) can thus probably be reconciled
through the age-metallicity degeneracy.

Therefore, despite having failed to find the same parameter ranges
when using spectral line strengths and colours, we do feel that we
have found plausible parameters that reproduce the observed radial
colour profiles of UV upturn and UV-weak galaxies. Furthermore, it
is important to reproduce the UV relations for internal properties
using the stellar population synthesis models of \citet{yi03}, which
are specifically designed to explain the UV upturn phenomenon.
Notably, our models reproduce the range of UV strengths among
elliptical galaxies by varying the metallicity. In the context of
the \citet{yi03} models, the UV upturn is a stellar phenomenon
associated with the metal-rich (high \mgb\ line strength) population
of early-type galaxies, and the UV-bright central regions of giant
elliptical galaxies (e.g. NGC 4552 and 4278) are enhanced in metals
by roughly 60\% compared to the UV-weak parts.

%
%
\section{SUMMARY}
\label{sec:conclusions}



We have used ultraviolet photometry from \galex, ground-based
optical photometry from MDM and ground-based optical integral-field
spectroscopy from \sauron\ to study the UV--line strength and
UV--stellar population relations of 34 early-type galaxies from the
\sauron\ sample. Our main advantage in exploring these correlations
is that all photometric and spectroscopic measurement are derived
using the same ellipses and apertures.

First, we have found that passively-evolving galaxies show clear
correlations between the colours involving FUV (FUV$-V$ and
FUV$-$NUV) and the \mgb\ and \hb\ absorption line strength indices,
as well as the metallicity [Z/H], and that these correlations are
strongest for FUV$-$NUV. This result is supported by a
pixel-by-pixel analysis. This suggests that either Mg is a main
driver of the UV upturn or Mg and the UV upturn share the same
primary driver, and that the correlation between the UV colour and
metallicity is explained by the role of Mg in the metallicity
measurement. Likewise, the correlation of the UV colour with \hb\
can be explained by the metallicity effect on \hb.

In addition, we have also derived logarithmic internal colour, {\it
measured} line strength and  stellar population gradients for each
galaxy and again found a strong dependence of the FUV$-V$ and
FUV$-$NUV colours on the \mgb\ line strength, as well as on
metallicity. Moreover, the \mgb\ and [Z/H] global gradients with
respect to the UV colours (e.g. $\Delta$(\mgb\ )/$\Delta$(FUV$-$V))
are consistent with their respective local gradients, implying that
the global correlations also hold locally within galaxies.

Finally, if the UV strength is indeed governed by stellar population
properties, then the same models should be able to reproduce the UV
relations for both integrated properties and internal properties
(within each galaxy). From simple models based on UV$-$optical
colour fits of UV upturn and UV-weak galaxies, we have identified a
plausible range of parameters that reproduces the observed radial
colour profiles of each galaxy type. In these models, the UV upturn
is a stellar phenomenon associated with the metal-rich (high \mgb\
line strength) population of early-type galaxies, and the inner
parts (UV-bright regions) of giant elliptical galaxies are enhanced
in metals by roughly 60\% (e.g. NGC 4552 and 4278) compared to the
outer parts (UV-weak regions). We stress here that stars originating
the UV upturn phenomenon are detected in the central region of
galaxies (especially within R${\rm_e}$/4), and therefore they are
contributing $<$\,20\,\% of the total stellar light. This means that
the UV upturn phenomenon is naturally produced if the galaxy has a
small fraction of Mg enhanced populations. Assuming that the galaxy
has a 10\% metal-enhanced population, it cannot make a significant
change in the overall metallicity of the galaxy, but this population
can evolve to the FUV excess stars.

Without doubt, we now have much more information on the UV upturn
than a decade ago. However, there still are many puzzles. For
example, it is surprising that the $\alpha$-enhancement
([$\alpha$/Fe]) does not show a correlation with UV strength,
considering that \mgb\ shows a tight correlation with UV colours and
[Z/H] also shows a marginal correlation. Improvements over the work
presented in this paper will come from a more careful population
modelling analysis.

Our achievements here were mainly obtained through the
spatially-resolved analyses that have just recently become possible.
A critical weakness of our work however is the poor spatial
resolution of the UV data. Breakthroughs will be guaranteed when
higher spatial resolution (arcsec scale) FUV data become available,
hopefully in the near future.
%
%
%
\section*{Acknowledgments}
This work was supported by the National Research Foundation of Korea
through the Doyak grant (No. 20090078756), the SRC grant to the
Center for Galaxy Evolution Research and the KASI grant given to
SKY. MB acknowledges support from NASA through GALEX Guest
Investigator program GALEX GI 04-0000-0109. MB and SKY are grateful
to the Royal Society for an International Joint Project award
(2007/R2) supporting this work. The STFC Visitors grant to Oxford
also supported joint visits. JF-B acknowledges the support from the
Ram\'on y Cajal Program as well as grant AYA2010-21322-C03-02 by the
Spanish Ministry of Science and Innovation. GALEX is operated for
NASA by the California Institute of Technology under NASA contract
NAS5-98034. Photometric data were also obtained using the 1.3m
McGraw-Hill Telescope of the MDM Observatory. Part of this work is
based on data obtained from the ESO/ST-ECF Science Archive Facility.
This project made use of the HyperLeda database
(http://leda.univ-lyon1.fr) and the NASA/IPAC Extragalactic Database
(NED) which is operated by the Jet Propulsion Laboratory, California
Institute of Technology, under contract with the National
Aeronautics and Space Administration.
%
%

%
%
\end{document}